\documentclass[aps,showpacs,prd,twocolumn,nofootinbib,nobibnotes]{revtex4-1}
\usepackage{graphicx}
\usepackage{amssymb}
\usepackage{amsmath}
\usepackage{color}
\usepackage{float}
\usepackage{ulem}
\usepackage{accents}
\usepackage{graphicx}
\usepackage{graphicx}
\usepackage{amsfonts}
\usepackage[colorlinks=true,
pdfstartview=FitV,linkcolor=blue,
citecolor=blue,urlcolor=blue,breaklinks=true]{hyperref}
\usepackage{array}
\usepackage{float}
\usepackage{placeins}
\usepackage[dvipsnames]{xcolor}
\usepackage{csquotes}
\usepackage{bbold}
\usepackage{units}
\usepackage{fixmath}
\usepackage{cancel}

\newcolumntype{C}[1]{>{\centering\arraybackslash}m{#1}}
\renewcommand{\eqref}[1]{\mbox{Eq.~(\ref{#1})}}

\definecolor{ForestForestGreen}{rgb}{0.13,0.55,0.13}
\definecolor{ForestGreen}{rgb}{0.13,0.55,0.13}

\begin{document}

\title{Optical properties and energy propagation in a dielectric medium \\ supporting magnetic current}

\author{Pedro D. S. Silva$^{a}$}
\email{pedro.dss@ufma.br} \email{pdiego.10@hotmail.com}
\author{M. J. Neves$^{b}$}\email{mariojr@ufrrj.br}
\author{Manoel M. Ferreira Jr.$^{a,c}$}
\email{manojr.ufma@gmail.com}
\affiliation{$^a$Programa de P\'{o}s-graduaç\~{a}o em F\'{i}sica, Universidade Federal do Maranh\~{a}o, Campus Universit\'{a}rio do Bacanga, S\~{a}o Lu\'is (MA), 65080-805, Brazil}
\affiliation{$^{b}$Departamento de F\'{i}sica, Universidade Federal Rural do Rio de Janeiro,
BR 465-07, Serop\'edica (RJ), 23890-971, Brazil}

\affiliation{$^c$Departamento de F\'{\i}sica, Universidade Federal do Maranh\~{a}o,
Campus Universit\'{a}rio do Bacanga, S\~{a}o Lu\'is (MA), 65080-805, Brazil}

\date{\today}

\begin{abstract}

We examine a dielectric medium supporting a magnetic current in connection with optical properties and energy propagation. The dispersion relations, propagating modes, and some optical effects are examined for isotropic and anisotropic magnetic conductivity tensors, with the latter ones implying nonreciprocal permittivities. The eigenvalues of the effective permittivity are carried out and associated with the optical symmetries (uniaxial, biaxial) and the subjacent crystal systems. Aspects of electromagnetic energy transport for such systems are also discussed, with the group and energy velocities being presented and carried out for all the particular cases addressed. Our results suggest that a dielectric supporting magnetic current breaks the usual equivalence between group velocity and energy velocity that holds in a nonabsorbing medium, while establishing the equivalence between them in an effective absorbing scenario, two unexpected behaviors.

\end{abstract}

\pacs{41.20.Jb, 78.20.Ci, 78.20.Fm}
\keywords{Electromagnetic wave propagation; Optical constants;
Magneto-optical effects; Birefringence}

\maketitle

\section{Introduction}
A magnetic linear current law, as ${\bf J}={\sigma}_{B}{\bf B}$, appears in some systems due to an asymmetry
between the number density of left- and right-handed chiral fermions as a macroscopic manifestation
of the quantum Chiral Magnetic Effect (CME) \cite{Kharzeev1}. Such an effect has been extensively investigated
in physics, passing in several distinct scenarios, such as quark-gluon plasmas with a chiral chemical potential
under the influence of an external magnetic field \cite{Fukushima,Gabriele}, cosmic magnetic fields in the early Universe \cite{Schober,Vilenkin},  Cosmology \cite{Maxim}, neutron stars \cite{Leite, Dvornikov}, and electroweak interactions \cite{Maxim1, Maxim2}. In condensed-matter systems, the CME is connected with the physics of Weyl semimetals (WSMs) \cite{Burkov}, where the massless fermions acquire a drift velocity along
the magnetic field, whose direction is given by their chirality. Opposite chirality implies opposite velocities, creating a chiral-fermion imbalance that is proportional to the chiral magnetic current. In WSMs, the chiral current may be different from the usual CME linear relation, ${\bf J}={\sigma}_{B}{\bf B}$, when parallel electric and magnetic fields are applied, yielding a current of the type ${\bf J}={\sigma}({\bf E} \cdot {\bf B}){\bf B}$, which corresponds to a conductivity effectively proportional to $B^{2}$ \cite{Li,Xiaochun-Huang,Barnes}. WSMs and CME have been examined in several scenarios and respects, considering the absence of Weyl nodes \cite{Chang}, anisotropic effects stemming from tilted Weyl cones \cite{Wurff}, the CME and anomalous transport in Weyl semimetals \cite{Landsteiner}, quantum oscillations
arising from the CME \cite{Kaushik}, computation of the electromagnetic fields produced by an electric charge near a topological Weyl semimetal with
two Weyl nodes \cite{Ruiz}, the chiral superconductivity \cite{Kharzeev2}, wave-packet scattering connected with transport properties \cite{Qing-Dong}, and evaluation of the Goos-Hänchen lateral shift for reflection at a Weyl semimetal interface \cite{Qing-Dong2}.
The parity-odd Maxwell-Carroll-Field-Jackiw model (MCFJ) \cite{CFJ1} is a Lorentz-violating electrodynamics characterized by the term,
	\begin{align}
	\mathrm{{\mathcal{L}_{CFJ}}}=-\frac{1}{4}\epsilon^{\mu\nu\alpha\beta}\left( k_{AF}\right)  _{\mu}A_{\nu}F_{\alpha
		\beta}, \label{MCFJMATTER}
\end{align}
where $\left(k_{AF}\right)_{\mu}$ is the 4-vector background which violates Lorentz symmetry. This \textit{CPT}-odd U(1) model is a piece of the Standard Model Extension (SME) \cite{Colladay}, proposed to examine and to constrain the possibility of Lorentz violation in nature. The MCFJ theory is a proper theoretical framework to address the CME in an effective classical way \cite{Qiu} since it provides a modified Amp\`ere's law including the magnetic current, $\mathbf{J}_{B}=k_{AF}^{0}\mathbf{B}$, where the zero component $k_{AF}^{0}$ works as the magnetic conductivity. The MCFJ model is also a possible version of the axion coupling \cite{Wilczek}, $\theta(\mathbf{E}\cdot \mathbf{B)}$, when the axion field presents constant derivative, $\left( k_{AF}\right)_{\mu}=\partial_{\mu}\theta$. Axion theories constitute a relevant topical research area actually \cite{KDeng, Sekine, Tobar}. A classical description of wave propagation, refractive indices, and optical effects, in the context of a matter medium governed by the  MCFJ electrodynamics, was discussed in Ref. \cite{Pedroo}, which also addresses higher-order derivative terms. Very recently, the MCFJ model has been connected with the London equation in parity-odd \cite{Stalhammar} and time-violating Weyl superconductors \cite{Shyta}. The MCFJ electrodynamics was much examined, including studies with radioactive corrections \cite{CFJ4}, topological defects solutions \cite{CFJ5}, supersymmetric generalizations \cite{CFJ6}, classical solutions with boundaries \cite{Borges}, classical dipole radiation \cite{AltschulR}, quantum aspects and unitarity analysis \cite{CFJ7}. Lorentz-violation in continuous media ~\cite{Bailey,Gomez} has been a focus of attention in the latest years due to its potential to describe interesting effects of the phenomenology of new materials, such as Weyl semimetals \cite{Marco}.

In a recent work \cite{Pedro1}, it was examined the classical wave propagation in a conventional dielectric medium, $\mathbf{D}={\epsilon }\,\mathbf{E}$, $\mathbf{B}={\mu }\,\mathbf{H}$, in the presence of a magnetic density current, $J^i= \sigma^B_{ij} B^j$, in which $\sigma^B_{ij}$ is the magnetic conductivity. This study was first performed for the case of an isotropic current, which describes the CME observed in Weyl semimetals,
$J_{\mathrm{CME}}^{i}={\sigma}^{B}\,B^{i}$, with the attainment of the refractive indices and electric fields for the propagating modes. Configurations
with the anisotropic conductivity tensor were also investigated, with some interesting results and repercussions in the literature ~\cite{Kaushik1}.
As well known, the electromagnetic energy balance in a continuous medium is stated by the Poynting's theorem \cite{Jackson,Zangwill, Landau},
\begin{eqnarray}
\nabla \cdot {\bf S} + \frac{\partial u_{EM}}{\partial t}=-\,{\bf J}\cdot{\bf E} \; ,
\end{eqnarray}
in which the Poynting vector is defined by
\begin{equation}
{\bf S}={\bf E}\times{\bf H}^{\ast} \; ,
\end{equation}
and $u_{EM}$ represents the electromagnetic energy density stored in the field. For an isotropic dispersive media such an energy density is given by \cite{Zangwill, Jackson}
\begin{eqnarray}\label{complex-group-velocity-isotropic-28}
u_{EM} &= \frac{1}{2} \frac{\partial [\omega \epsilon' ]}{\partial \omega} \left( {\bf{E}} \cdot {\bf{E}}^{*} \right) + \frac{1}{2}\frac{\partial [\omega \mu' ]}{\partial \omega} \left( {\bf{H}} \cdot {\bf{H}}^{*} \right) \; ,
\end{eqnarray}
where $\epsilon'$ and $\mu'$ are the real pieces of the permittivity and permeability, {\it i.e.},
\begin{align}
	\epsilon(\omega)=\epsilon^{\prime}(\omega)+i\epsilon^{\prime\prime}(\omega), \quad
	\mu(\omega)=\mu^{\prime}(\omega)+i\mu^{\prime\prime}(\omega) \; ,
\end{align}
also endowed with imaginary pieces, $\epsilon^{\prime\prime}(\omega)$, $\mu^{\prime\prime}(\omega)$, in the general situations.
For an electric dispersive medium ruled by the isotropic constitutive relations,
\begin{align}\label{constitutive-relations-Drude-1}
	{\bf{D}} &= \epsilon (\omega) \, {\bf{E}} \; , \quad {\bf{H}} =\mu^{-1} \, {\bf{B}} \, ,
\end{align}
whose Poynting vector and energy density are
\begin{subequations}
\begin{eqnarray}
{\bf S}\!&=&\!\frac{1}{2\mu'}\left({\bf E}\times{\bf B}^{\ast}\right) \; ,
\label{complex-group-velocity-isotropic-28}
\\
u_{EM} \!&=&\! \frac{1}{2} \frac{\partial [\omega \epsilon' ]}{\partial \omega} \left( {\bf{E}} \cdot {\bf{E}}^{*} \right) + \frac{1}{2\mu'} \left( {\bf{B}} \cdot {\bf{B}}^{*} \right) \; ,
\end{eqnarray}
\end{subequations}

where the fields, in general, may be complex,
\begin{equation}
 {\bf E}={\bf E}^{\prime} + i \, {\bf E}^{\prime\prime}, \quad {\bf B}={\bf B}^{\prime} + i \, {\bf B}^{\prime\prime}.
\end{equation}
Considering a plane-wave \textit{ansatz} for the fields,
\begin{subequations}
\begin{align}
\mathbf{(E,D)}=(\mathbf{E}_{0},\mathbf{D}_{0}) \, e^{\mathrm{i}(\mathbf{k}\cdot\mathbf{r}-\omega t)} \; , 	\label{plane1a}
\displaybreak[0]\\
\mathbf{(B,H)}=(\mathbf{B}_{0},\mathbf{H}_{0}) \, e^{\mathrm{i}(\mathbf{k}\cdot\mathbf{r}-\omega t)} \; ,
\label{plane1b}
\end{align}
\end{subequations}
where $\omega$ is the harmonic frequency, and the ${\bf k}$ is the wave vector, the Faraday's law reads $\omega \, {\bf B} =  {\bf k} \times {\bf E}$.
In the presence of absorption, the wave vector is complex,
\begin{equation}
{\bf k}={\bf k}^{\prime} + i \, {\bf k}^{\prime\prime},
\end{equation}
so that the Poynting vector and energy density reads
\begin{subequations}
\begin{align}\label{timeaverageSuEM1}
 \left\langle {\bf{S}} \right\rangle &=  \frac{{\bf{E}}^{2}}{2\omega\mu'} \, {\bf k}^{\prime}   \; ,
\\
\left\langle u_{EM} \right\rangle &= \frac{1}{4} \left[ \epsilon' + \omega \, \frac{ \partial \epsilon'}{\partial \omega} + \frac{ |{\bf{k}}|^{2}}{\mu \omega^{2}} \right] {\bf{E}}^{2} \; ,
\label{timeaverageSuEM2}
\end{align}
\end{subequations}
which holds for the case of total transversal modes, that is,
\begin{subequations}
\begin{align}
			\label{complex-group-velocity-isotropic-54A}
		({\bf{k}}' \cdot {\bf{E}}') = ({\bf{k}}'' \cdot {\bf{E}}'')=0, \\
		({\bf{k}}' \cdot {\bf{E}}'') =({\bf{k}}'' \cdot {\bf{E}}')=0. \label{complex-group-velocity-isotropic-54B}
	\end{align}
\end{subequations}
In a dispersive and non-absorbing medium ($\epsilon^{\prime\prime}=0$), the electromagnetic signal propagation is governed by the group velocity,
\begin{align}\label{complex-group-velocity-isotropic-26}
{\bf{v}}_{g} &= \frac{d \omega}{d k}  \hat{\bf{k}}  \; ,
\end{align}
which is a real quantity. For a dispersive and absorbing medium ($\epsilon^{\prime\prime}\neq0$), however, it becomes a complex velocity,
	\begin{align}\label{complex-group-velocity}
	{\bf{v}}_{g} &= \mathrm{Re}[d\omega/dk] + i  \mathrm{Im}[d\omega/dk]  \; \hat{\bf{k}} ,
	\end{align}
whose physical interpretation is usually unclear \cite{Neufeld}. Even in this scenario, the real and imaginary pieces of (\ref{complex-group-velocity}) may find physical interpretation, being useful to address aspects of Gaussian wave packets propagating in an absorbing medium \cite{Garrett,Connor}. Indeed, in the saddle point approximation the quantity $d\omega/d[Re(k)]$ yields the velocity of packet peak spatial maximum, $x_M$, while the imaginary part shifts the central wave number, $k_c$, which is in general no longer conserved \cite{Sonnenschein,ChenPRA}.

The failure of the complex group velocity (\ref{complex-group-velocity}) for describing energy transport in absorbing media was first noticed by Brillouin \cite{Brillouin}, leading to the definition of the energy velocity \cite{Loudon,Sherman,Davidovich,Ruppin}, a ratio between the averaged Poynting vector and the averaged energy density,
\begin{align}
\label{complex-group-velocity-isotropic-26}
{\bf{v}}_{E} &= \frac{ \left\langle {\bf{S}} \right\rangle} {\left\langle u_{EM} \right \rangle} \; ,
\end{align}
that also includes the dissipation terms. For an absorbing medium supporting transversal modes, the time averages (\ref{timeaverageSuEM1}) and (\ref{timeaverageSuEM2}) yield the energy velocity,
\begin{align}\label{vE}
{\bf{v}}_{E} =  \frac{ \displaystyle {{\bf{k}}'}/{(\mu\omega)} } {\displaystyle \frac{\epsilon'}{2} + \frac{\omega}{2} \frac{\partial \epsilon'}{\partial \omega} +  |{\bf{k}}|^{2}/{(2\mu\omega^{2})}} \; ,
\end{align}
which depends on the permittivity and the dispersion relation of the medium enclosing the absorption contribution (inside $|{\bf{k}}|^{2}$). For a non-absorbing medium, the group and energy velocities are equal, 	${\bf{v}}_{g} ={\bf{v}}_{E}$, while for an absorbing medium, such equality does not hold anymore,
\begin{align}\label{complex-group-velocity-isotropic-26}
	{\bf{v}}_{g} \neq {\bf{v}}_{E}=\frac{ \left\langle {\bf{S}} \right\rangle} {\left\langle u_{EM} \right \rangle} \; .
\end{align}

In lossy scenarios, where the group velocity becomes complex and different from ${\bf{v}}_{E}$, the question of physical velocity that can be measured arises. The time interval for a signal to reach its destination in a medium is called time delay, being used to define the signal velocity in dispersive and lossy media \cite{Oughstun1988}, which is a physical \cite{Oughstun1989} and measured quantity \cite{Brunner}. The energy velocity coincides with the signal velocity in certain frequency ranges \cite{Oughstun1994}, therefore representing the physical quantity measured by experimental devices \cite{Brunner, Centini} for these situations. Furthermore, reading from exotic group velocities information pieces about the signal velocity is also a relevant topic \cite{Peatross}.

Furthermore, it is known the dependence of the energy velocity on the way one evaluates the energy density and the dissipated energy density in a lossy medium, since there exist different forms to do it \cite{Cui,Semchenko}. In this sense, attempts to determine the correct expressions of storage energy and dissipated energy densities have been developed \cite{Chen21,Chen22}. Electromagnetic energy propagation and energy storage in a dispersive and absorbing hyperbolic metamaterial (HMM) have also been recently examined using the concept of energy velocity as far as its equality to the group velocity in a lossless HMM \cite{Moradi}.

 Connections between group and energy velocities were also explored in dissipative dynamical systems \cite{Gerasik}. Part of this procedure was used here to investigate the group and energy propagation in a chiral dielectric medium endowed with magnetic conductivity, which is still an open issue. Such a study is the main topic examined in the present work, which also complements the results attained in Ref. \cite{Pedro1}. In this context, we point out that a dielectric medium ($\epsilon^{\prime} \neq 0$, $\epsilon^{\prime\prime}=0$), in the presence of isotropic magnetic conductivity, presents real refractive indices and no absorbing behavior, see \eqref{n23-0-1}. Such a non-absorbing medium, however, possesses group velocity distinct from the energy velocity, an unexpected result stemming from the presence of magnetic conductivity. For an antisymmetric conductivity tensor, the opposite unexpected behavior is reported. Indeed, in this case, the refractive index is complex, there existing absorption, but it holds ${\bf{v}}_{g} ={\bf{v}}_{E}$.

This work is outlined as follows: Sec.~\ref{section2}, we review basic aspects of chiral electrodynamics in matter, examining the propagating modes, some optical effects, and also discussing features of parity and reciprocity. For each configuration of the conductivity tensor, the structure and eigenvalues of the permittivity are related to optical symmetries (uniaxial, biaxial) and possible crystal systems. In Sec.~\ref{section3} we discuss aspects of the energy stored in the electromagnetic field and energy flux by electromagnetic waves. By introducing the chiral magnetic current $J_{i} = \sigma_{ij}^{B} B_{j}$, we study the cases in which the magnetic conductivity tensor $\sigma^{B}_{ij}$ is diagonal (isotropic), antisymmetric, and symmetric. We investigate the group velocity and the energy propagation velocity for these particular scenarios. The group velocity and energy velocity are calculated in all these cases. We compare the scenarios of an absorbing and nonabsorbing medium with the results of the literature. In Sec.~\ref{conclusions} we summarize our results, also remarking aspects of plasmon-polaritons possible solutions and the magnitude of the magnetic current in physical systems.

%

%

%
%$\hbar=c=1$  with $4 \pi \epsilon_0 = 1$, and the Minkowski
%metric $\eta^{\mu\nu}=\mbox{diag}(+1,-1,-1,-1)$. The electric and magnetic fields have squared-energy mass dimension in which the conversion
%of Volt/m and Tesla (T) to the natural system is as follows: $1 \, \mbox{Volt/m}=2.27 \times 10^{-24} \, \mbox{GeV}^2$ and $1 \, \mbox{T} =  6.8 \times %10^{-16} \, \mbox{GeV}^2$, respectively.
%

%
\section{Electrodynamics with a magnetic conductivity and optical effects}
\label{section2}
In this section, we review some basic aspects of a classical electrodynamics in the presence of a magnetic current density, as previously developed in Ref. \cite{Pedro1}, also examining the correspondent optical effects in accordance with general procedures for anisotropic crystal media \cite{Ramachandran,Agranovich,Agranovich2}. We take as a starting point, the Maxwell equations for a linear, homogeneous, and isotropic medium are
\begin{subequations}
\begin{eqnarray}
\nabla\cdot{\bf D} \!&=&\! \rho \; ,
\hspace{0.3cm}  \hspace{0.3cm}
\nabla\times{\bf E}+\frac{\partial{\bf B}}{\partial t} = {\bf 0} \; ,
\label{eqdivD}
\\
\nabla\cdot{\bf B} \!&=&\!0 \; ,
\hspace{0.3cm}  \hspace{0.3cm}
\nabla\times{\bf H}-\frac{\partial{\bf D}}{\partial t} = {\bf J} \; ,
\label{eqdivB}
\end{eqnarray}
\end{subequations}
where $\rho$ and ${\bf J}$ are classical sources of charges and currents. Using the constitutive relations (\ref{constitutive-relations-Drude-1}), the magnetic current ${\bf J}={\sigma}_{B}\,{\bf B}$, plane-wave solutions \textit{ansatz} (\ref{plane1a}) and (\ref{plane1b}), and the Faraday's law,
the Ampere-Maxwell's law (\ref{eqdivB}) can be written as
\begin{align}
%{\bf{k}}\times {\bf{B}} +  \mu \epsilon \omega {\bf{E}} + \mathrm{i} \mu \sigma_{B} {\bf{B}}  =0\,,
%\label{mod-ampere-2-2B} \displaybreak[0]\\
{\bf{k}}\times {\bf{k}}\times {\bf{E}} +  \mu \, \epsilon \, \omega^2 \, {\bf{E}} + \mathrm{i} \, \mu \, \sigma_{B} \, {\bf{k}}\times {\bf{E}} = 0 \; .
\label{Ampere1bb}
\end{align}
For an anisotropic medium, the magnetic current,
\begin{equation}
{J}^{i}={\sigma}_{ij}^{B} \, {B}^{j} \; ,
\label{JCMEij}
\end{equation}
is written in terms of the conductivity tensor, $\sigma^{B}_{ij}$, depends on the material properties.
With it, \eqref{Ampere1bb} reads
\begin{equation}
({\bf{k}}\times {\bf{k}}\times {\bf{E}})_{i} +  \mu \, \epsilon \, \omega^2 \, {{E}_{i}}
+ \mathrm{i} \, \mu \, {\sigma}_{ij}^{B} ({\bf{k}} \times {\bf{E}})_{j}  =0\,.
\label{Ampere1b}
\end{equation}

The wave equation for the electric field amplitude is
\begin{align}\label{EqwaveE0}
\left[ \, {\bf k}^2 \, \delta_{ij} - k_{i}\,k_{j}-\omega^2 \mu \, \overline{\epsilon}_{ij}(\omega) \, \right] E_{j}=0 \; ,
\end{align}
where we have defined the permittivity tensor
\begin{eqnarray}
\overline{\epsilon}_{ij}(\omega)= \epsilon(\omega) \, \delta_{ij}
-\frac{i}{\omega^2} \, \sigma^{B}_{ia} \, \epsilon_{abj} \, k_{b} \; ,
\label{permittivityCME}
\end{eqnarray}
which in general is non-Hermitian, being thus associated with absorption effects \cite{Landau,Olivier}. Relation (\ref{EqwaveE0}) is also written as ${M}_{ij}E^{j}=0$, with
\begin{equation}
{M}_{ij}= {\bf k}^2 \, \delta_{ij} - k_{i}\,k_{j}-\omega^2 \mu \, \overline{\epsilon}_{ij}(\omega) \; .
\label{MijCME}
\end{equation}

The non-trivial solutions require that the determinant is null
\begin{eqnarray}
\label{Det}
	\det[{ M_{ij} }] = 0 \; ,
\end{eqnarray}
which provides the dispersion relations of the model. In the next sections, we consider materials whose constitutive relations are given by \eqref{constitutive-relations-Drude-1} and endowed with the magnetic conductivity (\ref{JCMEij}) parameterized through isotropic, symmetric
and antisymmetric configurations for the magnetic conductivity tensor $\sigma^{B}_{ij}$.
It is worth mentioning that the permittivity given in \eqref{permittivityCME} represents, in general, non-reciprocal media, since it does not satisfy the Onsager relations \cite{Landau,Agranovich,Agranovich2,Kong, Caloz, Buhmann}, namely
	\begin{eqnarray}
		{\epsilon}_{ij}(\omega, {\bf{k}})={\epsilon}_{ji}(\omega,-{\bf{k}}) \; ,
		\label{Onsager1}
	\end{eqnarray}
which constitutes a criterium to establish reciprocity in electromagnetic systems\footnote{These relations in electromagnetic systems are also known as Onsager-Casimir relations \cite{Buhmann}, being derived for the electromagnetic constitutive tensors considering the concept of reaction introduced by Rumsey \cite{Rumsey}.} \cite{Buhmann, Caloz, Rumsey}, in principle consistent with the symmetric permittivity approach \cite{Olivier}.
For continuous systems, the electric permittivity is generally given as an expansion of the wave vector, including the magnetic field, etc., providing a general dispersive behavior that may respect or not the reciprocity. Nonreciprocity is associated with interesting scenarios, such as the Faraday effect in magnetized plasmas, in such a way reciprocity or nonreciprocity can be stated in terms of the natural activity or magnetic optical activity of the medium \cite{Barron}, respectively. Nonreciprocity can be also achieved in nonlinear systems \cite{Roy}, metamaterials \cite{Sounas}, space-time modulated structures \cite{Hadad-Sounas}, moving media \cite{Wang}, non-hermitian electromagnetic response \cite{Silveirinha-nonreciprocity} and time reversal symmetry \cite{Siddharth}.
In accordance with the Onsager relation (\ref{Onsager1}), dielectric media endowed with the magnetic conductivity examined in the present work manifest nonreciprocity in terms of the off-diagonal components of the conductivity tensor, $\sigma^{B}_{ij}$. Indeed, the permittivities written for the off-diagonal magnetic conductivity break the symmetry of the Onsager relation and are non-Hermitian, yielding absorption factors that appear in terms of the imaginary pieces of the corresponding refractive indices.
Aspects of nonreciprocal electromagnetic response in Weyl semimetals (due to nonsymmetric permittivity) are a relevant topic \cite{Guo,Trepanier,Cote,Zhao,Hofmann}, with analog magnetoelectric properties to the ones of the present work. In the WSM electrodynamics, the constitutive relation for the displacement vector \cite{Guo} can be written as
\begin{align}\label{WSMconstitutive-relation1}
	{\bf{D}} &= \epsilon_{D} (\omega){\bf{E}} +\frac{i \lambda \, e^2}{2\pi^2 \hbar\omega} (-k_{AF}^{0}\bf{B}+{\bf{k}}_{AF} \times {\bf{E}}) \; ,
\end{align}
with $k_{AF}^{0}$ and ${\bf{k}}_{AF}$ being the components of the 4-vector that appears in Lagrangian (\ref{MCFJMATTER}). Such a constitutive relation yields the WSM modified permittivity \cite{Trepanier},
\begin{eqnarray}
	\overline{\epsilon}_{ij}(\omega)= \epsilon\delta_{ij}
	+\frac{\lambda e^2}{\pi\hbar}\frac{i}{\omega} \epsilon_{ijl} \left[(k_{AF})_l- \frac{k_{AF}^{0}k_l}{ \omega} \right] \; ,
	\label{WSMpermittivityCME}
\end{eqnarray}
endowed with parity violation (term containing $k_{AF}^{0}$) and time reversal violation (term containing ${\bf{k}}_{AF}$). It is worth to note that the permittivity (\ref{permittivityCME}) is not symmetric (and does not respect the Onsager relation), implying violation of reciprocity. Thus, a WSM is nonreciprocal medium endowed with optical activity (birefringence) in the absence of an external magnetic field \cite{Cote}. One of the remarkable properties of the WSM is the giant optical nonreciprocity \cite{Guo, Zhao}, a property measured in terms of the asymmetry of the permittivity tensor, ${\gamma} = {|\epsilon_{ij} - \epsilon_{ji}|}/{|\epsilon_{ij} + \epsilon_{ji}|}$, since the reciprocity is associated with the permittivity symmetry. Here, $|... |$ designates the matrix norm. The $\gamma$ factor, in its numerator, captures the contribution of the $\epsilon$ asymmetric terms. These developments reported in the literature state that the nonreciprocity is a reality for the WSM ruled by the axion electrodynamics. Nonreciprocal repercussions are known in some scenarios. In fact, in photonics it is used to examine the violation of Kirchhoff's law of the thermal radiation in the absence of an external magnetic field, including conditions to maximize the nonreciprocal permittivity coefficients \cite{Zhao}. The attainment of nonreciprocal plasmon surface modes in WSM was also accomplished in connection with technological applications in nanoplasmonics \cite{Hofmann}. In the present work, in analogy to the WSM electrodynamics, we verify that the nonreciprocity is also a reality for the dielectric permittivity (\ref{permittivityCME}).
 The components of the tensor $\sigma_{ij}^{B}$ may also be related to the symmetries of the known crystal systems, in principle reproduced by the permittivity tensor of the dielectric endowed with magnetic conductivity, given in Eq. (\ref{permittivityCME}). Further, it is well stated that the crystal symmetries imply specific permittivity tensor structures \cite{Amnon, Robert}, which characterize the type of optical symmetry (cubic, uniaxial, or biaxial) associated. In this sense, considering the crystal symmetries on the effective permittivity (\ref{permittivityCME}), one can achieve some restrictions on the components of the tensor $\sigma^{B}_{ij}$. In principle, the cubic structure is compatible with until seven non-null components for $\sigma^{B}_{ij}$ (in the principal axes optical system);  for uniaxial materials, the tensor $\sigma^{B}_{ij}$ can have until eight non-null components; for biaxial crystals, the tensor $\sigma^{B}_{ij}$ can have until nine non-null components. Details about it are discussed in Appendix \ref{symmetry-restriction-appendix}.

\subsection{\label{particular2}Isotropic diagonal chiral conductivity}
\label{particular2}
As a first situation, we review the case of an isotropic magnetic conductivity, that is, $\sigma^{B}_{ij}=\Sigma\,\delta_{ij}$, where $\Sigma$ is a real parameter, which replaced in \eqref{permittivityCME}, yields
\begin{equation}
{\bar{\epsilon}}_{ij}(\omega )=\epsilon \, {\delta }_{ij}-{\frac{{\mathrm{i}\Sigma }}{{\omega }^{2}}} \, \epsilon_{ijb} \, k_{b} \, ,
\label{eq61a}
\end{equation}
which constitutes a Hermitian permittivity, entailing the absence of dissipation, as seen in the refractive indices to be obtained below.

Considering the behavior of the general constitutive relation, $J^{i} =\sigma^{B}_{ij} B^{j}$, under the inversion transformation, one finds that $\sigma^{B}_{ij}$ is odd under parity \cite{Kharzeev2,Pedro1}. Therefore, the conductivity $\Sigma$ should be a pseudo-scalar for assuring the permittivity (\ref{eq61a}) parity-invariant.

For this case, the matrix $M_{ij}$ (\ref{MijCME}), using ${\bf{n}}={\bf{k}}/\omega$, reads
\begin{equation}
\label{eq62}
[M_{ij}]=\mathcal{N}+\left(\begin{array}{ccc}
0 & \mathrm{i}\mu n_{3}\frac{\Sigma}{\omega} & -\mathrm{i}\mu n_{2}\frac{\Sigma}{\omega}\\
\\
-\mathrm{i}\mu\frac{n_{3}\Sigma}{\omega} & 0 & \mathrm{i}\mu n_{1}\frac{\Sigma}{\omega}\\
\\
\mathrm{i}\mu n_{2}\frac{\Sigma}{\omega} & -\mathrm{i}\mu n_{1}\frac{\Sigma}{\omega} & 0
\end{array}\right),
\end{equation}
where
\begin{equation}
\label{eq66-2}
\mathcal{N}=\begin{pmatrix}
		n_2^2+n_3^2-\mu\epsilon & -n_1n_2 & -n_1n_3 \\
		-n_1n_2 & n_1^2+n_3^2-\mu\epsilon & -n_2n_3 \\
		-n_1n_3 & -n_2n_3 & n_1^2+n_2^2-\mu\epsilon \\
	\end{pmatrix}\,.
\end{equation}
Requiring the null determinant, we obtain
\begin{equation}\label{eq63}
n_{\pm}^{2}=\mu\epsilon+2\left(\frac{\mu\Sigma}{2\omega}\right)^2 \pm \frac{\mu\Sigma}{\omega} \,
\sqrt{\mu\epsilon+\left(\frac{\mu\Sigma}{2\omega}\right)^2} \; ,
\end{equation}
so that the refractive indices are
\begin{subequations}
\label{n23-0-1}
\begin{eqnarray}
n_{\pm}&=&\sqrt{\mu\epsilon+\left(\frac{\mu\Sigma}{2\omega}\right)^2}\pm \frac{\mu\Sigma}{2\omega} \; ,
\label{n23-1A}
\\
\tilde{n}_{\pm}&=&-\sqrt{\mu\epsilon+\left(\frac{\mu\Sigma}{2\omega}\right)^2}\pm \frac{\mu\Sigma}{2\omega} \; ,
\label{n23-1B}
\end{eqnarray}
\end{subequations}
corresponding to four distinct real refractive indices. The two positive ones, $n_{\pm}$, were already examined in Ref. \cite{Pedro1} and define a dispersive non-absorbing medium compatible with the birefringence. We also note that there is no absorption (the indices are real), so the chiral conductivity does not imply a conducting behavior for the dielectric medium. The conducting behavior is assured only when it is defined simultaneously with the Ohmic conductivity $(\sigma \neq 0,\sigma ^{B}\neq 0) $. The propagation modes for the indices (\ref{n23-1A}), obtained from the relation $M_{ij}E_{j}=0$ \cite{Pedro1}, are
\begin{equation}
\label{eigenvector2}
\mathbf{E}_{\pm}=\frac{1}{n\sqrt{2(n_1^2+n_3^2)}}
\begin{pmatrix}
n n_3 \mp \mathrm{i}n_1 n_2 \\
\pm\mathrm{i}(n_1^2+n_3^2) \\
\mp\mathrm{i}n_2 n_3-n n_1 \\
\end{pmatrix} \; .
\end{equation}
We can write the modes for some propagation directions. As an initial case, we take a wave propagating at the ${\cal Z}$-axis,
$\mathbf{n}=(0,0,n_{3})$, for which Eq. (\ref{eigenvector2}) yields
\begin{equation}
	\label{eq:polarizations-isotropic}
	\mathbf{E}_{\pm}=\frac{1}{\sqrt{2}}\begin{pmatrix}
		1 \\
		\pm\mathrm{i} \\
		0 \\
	\end{pmatrix}\,.
\end{equation}
representing the left-handed ($L$) and right-handed ($R$) polarized electromagnetic waves, corresponding to $\mathbf{E}_{+}$ and $\mathbf{E}_{-}$, respectively. As already mentioned, a known consequence of the optical activity of a medium is linear birefringence, occurring when two circularly polarized modes of opposite chiralities, with refractive indices $n_{+}$ and $n_{-}$, respectively, have different phase velocities, $c/n_{+}$ and $c/n_{-}$. This property implies a rotation of the polarization plane of a linearly polarized wave. It is quantified by the specific rotatory power $\delta$, defined as
\begin{equation}
\label{eq:rotatory-power1}
\delta=-\frac{\omega}{2}[\mathrm{Re}(n_{+})-\mathrm{Re}(n_{-})] \; ,
\end{equation}
which measures the rotation of the oscillation plane of linearly polarized light per unit traversed length in the medium. Here, $n_{+}$ and $n_{-}$ are associated with left and right-handed circularly polarized waves, respectively. For the refractive indices (\ref{n23-1A}) and (\ref{n23-1B}),
the rotatory power is frequency-independent, depending only on the chiral magnetic
conductivity $\Sigma$, namely,
\begin{equation}\label{eq:rotatory-power}
\delta=-\frac{\mu\Sigma}{2} \; .
\end{equation}
For the wave propagation on ${\cal Z}$-direction, the permittivity tensor (\ref{eq61a}) is set by the matrix 
\begin{eqnarray}
\left[ \bar{\epsilon} \right]=
\left(
\begin{array}{ccc}
\epsilon & -\frac{ik_z}{\omega} \, \Sigma & 0 \\
\frac{ik_z}{\omega} \, \Sigma & \epsilon & 0 \\
0 & 0 & \epsilon \\
\end{array}
\right) \; , \label{PSigma1}
\end{eqnarray}
which exhibits three distinct eigenvalues: $\epsilon, \epsilon_{\pm}=\epsilon\pm \Sigma k_{z} /\omega$, playing the role of the principal permittivity values. This result shows that the material with permittivity (\ref{PSigma1}) may behave like a biaxial crystal (see the ref. \cite{Robert}) after reduced to its principal axes system. Biaxial systems, such as the one described in \eqref{PSigma1}, can also be recovered in chiral matter with anomalous Hall effect \cite{Qiu} described by the spacelike Maxwell-Carroll-Field-Jackiw electrodynamics \cite{Pedroo}.

For a general propagation direction, ${\bf{n}}={(k_1,k_2,k_3)}/\omega$, the permittivity tensor (\ref{eq61a}) reads
\begin{align}
	\left[ \bar{\epsilon}_{ij} \right] &= \begin{pmatrix}
		\epsilon && \frac{i\Sigma}{\omega^{2}} k_{3} && - \frac{i \Sigma}{\omega^{2}} k_{2} \\
		\\
		-\frac{i \Sigma}{\omega^{2}} k_{3} && \epsilon && \frac{i\Sigma}{\omega^{2}} k_{1} \\
		\\
		\frac{i \Sigma}{\omega^{2}} k _{2} && - \frac{i \Sigma}{\omega^{2}} k_{1} && \epsilon
	\end{pmatrix},  \label{crystal-symmetry-23}
\end{align}
also yielding three distinct principal values: $\epsilon$, $\epsilon_{+} = \epsilon + \frac{\Sigma}{\omega^{2}} k$,  $\epsilon_{-} = \epsilon - \frac{\Sigma}{\omega^{2}} k $, where $k=\sqrt{{\bf{k}}^{2}}$. Thus, such a tensor recovers the behavior of biaxial crystals (embracing triclinic, monoclinic, and orthorhombic systems).

\subsection{Nondiagonal antisymmetric conductivity}
\label{sec:nondiagonal-antisymmetric-configuration}

The magnetic conductivity tensor may be antisymmetric, ${\sigma }_{ij}^{B}=-{\sigma }_{ji}^{B}$, being parameterized as:
\begin{equation}\label{eq65}
{\sigma}_{ij}^{B}={\epsilon}_{ijk} \, b_{k} \; ,
\end{equation}

where $b_{k}=(b_{1},b_{2},b_{3})$ is a constant three-vector, and $\epsilon_{ijk}$ is the Levi-Civita symbol.
In this case, the permittivity (\ref{permittivityCME})
\begin{equation}
{\bar{\epsilon}}_{ij}(\omega)=\left[\epsilon - \,
\frac{\mathrm{i} ({\mathbf{k}\cdot \mathbf{b})}}{{\omega }^{2}}\right] {\delta }_{ij}
+\frac{\mathrm{i}}{{\omega}^{2}} \, k_{i} \, b_{j} \; ,  \label{eq66}
\end{equation}
implies nonreciprocity, since $\overline{\epsilon}_{ij}(\omega,k) \neq \overline{\epsilon}_{ji}(\omega,-k)$, and in conformity with the nonsymmetric behavior of $\bar{\epsilon}_{ij}$ \cite{Olivier, Guo}. Note this property is also shared by the WSM permittivity (\ref{WSMconstitutive-relation1}), as pointed out in Refs. \cite{Trepanier,Guo,Cote,Zhao}. Under the parity inversion, ${\bf k} \rightarrow -{\bf k}$, the permittivity tensor (\ref{eq66}) is kept invariant 
if ${\bf b}$ is a vector, {\it i.e.}, it transforms as ${\bf b} \rightarrow - {\bf b}$ (under parity). 

By replacing \eqref{eq66} in the matrix (\ref{MijCME}), we obtain:

\begin{align}
\label{eq66-1}
[&M_{ij}]=\mathcal{N} -\mathrm{i}\frac{\mu}{\omega} \nonumber \\
&\times\begin{pmatrix}
-(n_2b_2+n_3b_3) & n_1b_2 & n_1b_3 \\
n_2b_1 & -(n_1b_1+n_3b_3) & n_2b_3 \\
n_3b_1 & n_3b_2 & -(n_1b_1+n_2b_2) \\
\end{pmatrix}  ,
\end{align}

where
\begin{equation}\label{eq66-2}
\mathcal{N}=\begin{pmatrix}
n_2^2+n_3^2-\mu\epsilon & -n_1n_2 & -n_1n_3 \\
-n_1n_2 & n_1^2+n_3^2-\mu\epsilon & -n_2n_3 \\
-n_1n_3 & -n_2n_3 & n_1^2+n_2^2-\mu\epsilon \\
\end{pmatrix} \, .
\end{equation}
For $\det[M_{ij}]=0$, one obtains a doubly degenerate dispersion equation in $\mathbf{n}$,
\begin{equation}
\left[n^{2}+\mathrm{i}{\frac{\mu }{\omega }}\left(\mathbf{b}\cdot\mathbf{n}\right)
-\mu \epsilon\right]^{2}=0 \; ,
\label{eq67B}
\end{equation}
which provides the solution for the refractive index,
\begin{equation}\label{eq68}
%\begin{align}
n_{\pm}=\pm\sqrt{\mu\epsilon-\left( \frac{\mu}{2\omega}b\cos\theta  \right)^2}-\mathrm{i}\frac{\mu}{2\omega}b\cos\theta \, ,
%\label{eq67a}
%	\end{align}
\end{equation}
where $\mathbf{b}\cdot \mathbf{n}=b\,n\,\cos \theta $ with $b=|\mathbf{b}|$. For $\mu\epsilon \geq (\mu b\cos\theta/2\omega)^2$ and $\mu\epsilon\leq (\mu b\cos\theta/2\omega)^2$, the refractive index is complex and a pure imaginary, respectively.  The imaginary piece implies an anisotropic and dispersive absorption, ruled by a direction-dependent absorption coefficient, given by ${\tilde{\alpha}}=\mu \, b \, \cos \theta$. The real piece, when negative, corresponds to the negative refraction, that is usual in metamaterials.
The propagation modes are found for $\mathbf{n}$ along the $z$-axis and the vector $\mathbf{b}$ written as
\begin{equation}
\mathbf{n}=(0,0,n)\,,\quad \mathbf{b}=b\left(0,\sin\theta,\cos\theta\right)\equiv (0,b_2,b_3) \; ,
\label{COORD}
\end{equation}
where $\theta$ is the angle between $\mathbf{n}$ and $\mathbf{b}$. The matrix (\ref{eq66-1}) is
\begin{equation}
\lbrack M_{ij}]=\left(
\begin{array}{ccc}
n^{2}-\mu\epsilon + n \frac{ \mathrm{i}\mu  }{\omega} b_3 & 0 & 0 \\
0 & n^{2}-\mu\epsilon + n \frac{\mathrm{i}\mu }{\omega} b_3 & 0 \\
0 & - n \frac{\mathrm{i} \mu  }{\omega} b_2 & -\mu\epsilon \\
\end{array}
\right) \; ,
\label{MATRIX}
\end{equation}
with $\mathbf{n\cdot b}=n b_3$. For the condition $\mu\epsilon\geq (\mu b_{3}/2\omega)^2$, the indices (\ref{eq68}) possess a real piece. For the index $n_{+}$, the condition $M_{ij}E_j=0$ provides two mutually orthogonal propagation modes
\begin{equation}
\mathbf{E}_{\pm}^{(1)}=\frac{1}{\sqrt{2(1+Q^2)}}\begin{pmatrix}
\pm\sqrt{1+Q^2} \\
-1 \\
\mathrm{i} Q e^{\mathrm{i}\alpha}\\
\end{pmatrix} \; ,
\label{modesantisymmetric}
\end{equation}
while the index $n_{-}$ has associated,
\begin{equation}
\mathbf{E}_{\pm}^{(2)}=\frac{1}{\sqrt{2(1+Q^2)}}\begin{pmatrix}
\pm\sqrt{1+Q^2} \\
1 \\
-\mathrm{i} Q e^{\mathrm{-i}\alpha}     \\
\end{pmatrix}\,,
\label{modesantisymmetric1b}
\end{equation}
that fulfill the relations $\mathbf{E}_+^{(1)*} \cdot\mathbf{E}_-^{(1)}=0$, $\mathbf{E}_+^{(2)*} \cdot\mathbf{E}_-^{(2)}=0$, with $Q=b_2N/\epsilon \omega$.
Here, we parameterized the complex refractive index as $n=Ne^{\mathrm{i}\alpha}$, with $N=\sqrt{n^{*}n}=\sqrt{\mu\epsilon}$, and
\begin{equation}
\tan \alpha=\frac{\mu b_{3}/2\omega}{\sqrt{\mu\epsilon-\left(\mu/2\omega b_{3} \right)^2}} \; .
\end{equation}
The fields (\ref{modesantisymmetric}) set non orthogonal modes associated with $n_{+}$, while (\ref{modesantisymmetric1b}) stands for non orthogonal modes associated with $n_{-}$. If birefringence originates from two linearly polarized modes having different phase velocities, this property is not suitably characterized in terms of the usual rotatory power given by \eqref{eq:rotatory-power1}, since the latter is based on a decomposition of a linearly polarized mode into two circularly polarized ones of different chirality. For the case $b_{2}=0$, the solutions  (\ref{modesantisymmetric}) and (\ref{modesantisymmetric1b}) reduce to linearly polarized orthogonal modes,
\begin{equation}
\mathbf{E}_{\pm}^{(1)}=\frac{1}{\sqrt{2}}\begin{pmatrix}
\pm 1 \\
-1 \\
0
\end{pmatrix}, \quad
\mathbf{E}_{\pm}^{(2)}=\frac{1}{\sqrt{2}}\begin{pmatrix}
\pm 1 \\
1 \\
0
\end{pmatrix} \; .
\label{modesantisymmetric1c}
\end{equation}
In this case, the birefringence is measured in terms of the phase shift developed between the propagating modes (see Eq.~(8.32) in \cite{Hecht}):
\begin{equation}
\frac{\Delta}{l}=\frac{2\pi}{\lambda_0}\left[ \, \mathrm{Re}(n_{+})-\mathrm{Re}(n_{-}) \, \right] \; ,
\label{phase-shift1}
\end{equation}
where $\lambda_0$ is the wavelength of the electromagnetic radiation \textit{in vacuo}, and $l$ is the distance in which
the wave travels in the birefringent medium. Considering the refractive indices (\ref{eq68}), the phase shift per unit length is
\begin{equation}
\frac{\Delta}{l}=\frac{2\pi}{\lambda_0} \, \sqrt{\mu\epsilon-\left(\frac{\mu}{2\omega b_{3}} \right)^2} \; ,
\label{phase-shift2}
\end{equation}
also written as
\begin{align}
\frac{\Delta}{l}=\omega \, \sqrt{\mu\epsilon-\left( \frac{\mu}{2\omega b_{3}} \right)^2} \; . \label{birefringence-3}
\end{align}
As the refractive indices (\ref{eq68}) possess an imaginary piece, there is absorption for both modes (in equal magnitude), measured by an absorption coefficient ~\cite{Zangwill}, given by $\gamma=2\omega \, \mathrm{Im}(n)$, that is,
\begin{equation}
\gamma=\mu \, b_{3} \; .
\label{dichroism-2}
\end{equation}
In this case, by definition, there is no dichroism.
The permittivity tensor (\ref{eq66}) for the propagation on ${\cal Z}$-direction $k_i=(0,0,k)$, and $b_i=(0,b_2,b_3)$ is given by the matrix
\begin{eqnarray}
\left[ \epsilon \right]=
\left(
\begin{array}{ccc}
\epsilon-\frac{ik\,b_3}{\omega^2} & 0 & 0 \\
0 & \epsilon-\frac{ik\,b_3}{\omega^2} & 0 \\
0 & \frac{ik\,b_2}{\omega^2} & \epsilon \\
\end{array}
\right) \;, \label{PermTb}
\end{eqnarray}
whose diagonalization yields the two equal principal values: $\epsilon, \epsilon_{\pm}=\epsilon-ikb_{3}/\omega$. 
At this form, the permittivity tensor (\ref{PermTb}) recovers the behavior of an uniaxial crystal \cite{Robert}, belonging to the tetragonal, hexagonal, or trigonal systems. For a general propagation direction, ${\bf{n}}={(k_1,k_2,k_3)}/\omega$, and ${\bf{b}}$ vector, the permittivity (\ref{eq66}) also yields two different eigenvalues, $\epsilon, \epsilon- \frac{i}{\omega^{2}} ({\bf{k}}\cdot {\bf{b}})$, scenario compatible with uniaxial materials as well \cite{Amnon, Robert}. See Appendix \ref{symmetry-restriction-appendix} for more details.

\subsection{Nondiagonal symmetric conductivity tensor}
\label{sec:nondiagonal-symmetric-configuration}
Now we examine the case in which the magnetic conductivity is given by a traceless
symmetric tensor, in accordance with the following parametrization:
\begin{equation}
\sigma_{ij}^{B}=\frac{1}{2}\left(a_{i}\,c_{j}+a_{j}\,c_{i}\right) \; ,
\label{eq68-1}
\end{equation}
where $a_{i}$ and $c_{i}$ are the components of two orthogonal background
vectors $\mathbf{a}$ and $\mathbf{c}$, {\it i.e.}, 	$\mathbf{a} \cdot \mathbf{c}=0$,
such that $\sigma_{ii}^{B}=0$. Inserting \eqref{eq68-1} in the permittivity tensor (\ref{permittivityCME}), one obtains
\begin{equation}
\bar{\epsilon}_{ij}= \epsilon \, \delta_{ij}+{\frac{\mathrm{i}}{{2{\omega }^{2}}}}\left(
a_{i}\,c_{n}+a_{n}\,c_{i}\right) {\epsilon }_{nbj} \, k_{b} \; , \label{eq68-2}
\end{equation}
which also manifests nonreciprocity,  $\bar{\epsilon}_{ij}(\omega,k) \neq \bar{\epsilon}_{ji}(\omega,-k)$, and in accordance with Refs. \cite{Trepanier,Guo,Cote,Zhao}.
Taking into account that the tensor (\ref{eq68-1}) is P-odd, the permittivity (\ref{eq68-2}) preserves the parity symmetry in two situations: i) when ${\bf a}$ transforms as a pseudo-vector 
$({\bf a} \rightarrow {\bf a})$ and ${\bf c}$ as a vector; ii) when ${\bf a}$ is a vector and ${\bf c}$ a pseudo-vector.

The tensor (\ref{MijCME}) is explicitly represented by the following matrix:
\begin{widetext}
\begin{align}
\label{eq68-3}
[M_{ij}]=\mathcal{N}-\mathrm{i}\frac{\mu}{2\omega}\begin{pmatrix}
\epsilon_{11} & n_1(a_1c_3+a_3c_1)-2n_3a_1c_1 & -n_1(a_1c_2+a_2c_1)+2n_2a_1c_1 \\
-n_2(a_2c_3+a_3c_2)+2n_3a_2c_2 & \epsilon_{22} & n_2(a_2c_1+a_1c_2)-2n_1a_2c_2 \\
n_3(a_3c_2+a_2c_3)-2n_2a_3c_3 & -n_3(a_3c_1+a_1c_3)+2n_1a_3c_3 &\epsilon_{33} \\
\end{pmatrix} \; ,
\end{align}
\end{widetext}
where $\mathcal{N}$ is given by the \eqref{eq66-2}, and
\begin{subequations}
\begin{align}
\epsilon_{11}& =(a_{1}c_{2}+a_{2}c_{1})n_{3}-(a_{1}c_{3}+a_{3}c_{1})n_{2}\,,
\label{eq68-4} \displaybreak[0]\\[2ex]
\epsilon_{22}& =(a_{2}c_{3}+a_{3}c_{2})n_{1}-(a_{1}c_{2}+a_{2}c_{1})n_{3}\,,
\label{eq68-5} \displaybreak[0]\\[2ex]
\epsilon_{33}& =(a_{3}c_{1}+a_{1}c_{3})n_{2}-(a_{3}c_{2}+a_{2}c_{3})n_{1}\,.
\label{eq68-6}
\end{align}
\end{subequations}
The evaluation of $\det[M_{ij}]=0$ yields the forthcoming dispersion equation:
\begin{eqnarray}
\left[n^2-\mu\epsilon +\mathrm{i}\frac{\mu
}{2\omega}\mathbf{n}\cdot(\mathbf{a}\times \mathbf{c})\right]\times
\notag \\
\times\left[n^2-\mu\epsilon -\mathrm{i}\frac{\mu
}{2\omega}\mathbf{n}\cdot(\mathbf{a}\times \mathbf{c})\right]=0\,.
\label{eq68-8}
\end{eqnarray}
Using that $\mathbf{n}\cdot (\mathbf{{a}\times {c}})=n|\mathbf{a}||\mathbf{c}|\cos\varphi$, the
\eqref{eq68-8} yields the solutions
\begin{equation}
n_{\pm}=\sqrt{\mu\epsilon-\left(\frac{\mu}{{4\omega}}|\mathbf{a}||\mathbf{c}|\cos\varphi\right)^2} \pm\mathrm{i}\frac{\mu}{{4\omega}}|\mathbf{a}||\mathbf{c}|\cos\varphi\,.
\label{nac}
\end{equation}
The structure of the two refractive indices present dependence of the refractive index on an
angle between $\mathbf{k}$ and a three-vector ($\mathbf{b}$ for the antisymmetric case and $\mathbf{a}\times\mathbf{c}$ for the
current scenario).

In order to examine the propagating modes, we set the magnetic conductivity vectors of the symmetric case as
${\bf{a}}=(a_{1}, 0, a_{3} )$, ${\bf{c}} = (0, c_{2}, 0)$, such that one obtains
\begin{align}
{\bf{a}}\times {\bf{c}}= (-a_{3}c_{2}, 0, a_{1}c_{2} ) \; .
\label{symmetric-propagating-3}
\end{align}
For propagation along the ${\cal Z}$-axis, ${\bf{n}}=(0, 0, n)$, the matrix $M_{ij}$ (\ref{eq68-3}) is
\begin{align}
[M_{ij}] &= \begin{pmatrix}
n^{2}-\mu{\epsilon} - i \frac{\mu}{2\omega} a_{1}c_{2} n & 0 & 0 \\
0   &    n^{2}-\mu{\epsilon} + i \frac{\mu}{2\omega} a_{1}c_{2} & 0 \\
- i \frac{\mu}{2\omega} a_{3}c_{2} n  & 0 &  - \mu{\epsilon}
\end{pmatrix}  , \label{symmetric-propagating-4}
\end{align}
where ${\epsilon}$ is the complex electric permittivity. Evaluating $\mathrm{det}[M_{ij}]=0$, one finds the following dispersion relations:
\begin{align}
n^{2}-\mu{\epsilon}= \pm i \frac{\mu}{2\omega} \, a_{1} \, c_{2} \, n \; ,
\label{symmetric-propagating-5}
\end{align}
which is compatible with \eqref{eq68-8}.
We then write now the electric field polarizations satisfying $M_{ij}E_{j}=0$ associated with $n^{2}_{+}$ and $n^{2}_{-}$, respectively
\begin{align}
\hat{\bf{E}}_{+} = \frac{1}{\sqrt{1+|A|^2}}  \begin{pmatrix}
1 \\
0 \\
-i A
\end{pmatrix} \, ,
\quad
\hat{\bf{E}}_{-} = \begin{pmatrix}
	0\\
	1\\
	0
\end{pmatrix} \, ,
\label{symmetric-propagating-11}
\end{align}
where $A=a_{3}c_{2}n/(2\omega{\epsilon})$. We observe that $\hat{\bf{E}}_{\pm}$ represent the
linearly polarized vectors, with $\hat{\bf{E}}_{+}$ endowed with a longitudinal component. In the case we set $a_{3}=0$, the linearly polarized propagating modes become orthogonal, that is,
\begin{align}
	\hat{\bf{E}}_{+} =   \begin{pmatrix}
		1 \\
		0 \\
		0
	\end{pmatrix} \, ,
	\quad
	\hat{\bf{E}}_{-} = \begin{pmatrix}
		0\\
		1\\
		0
	\end{pmatrix} \, .
	\label{symmetric-propagating-11B}
\end{align}
If we consider refractive indices from (\ref{eq68-8}) with the positive real piece, the propagation turns out free of birefringence.
However, since the imaginary pieces are different, the propagating modes are absorbed in different degrees, which can be measured by the absorption difference per unit length,
\begin{equation}
\frac{\Delta_{d}}{l}=\frac{2\pi}{\lambda_0} \, \left[ \, \mathrm{Im}(n_{+})-\mathrm{Im}(n_{-}) \, \right]\,,
\label{phase-shift1}
\end{equation}
which for the indices (\ref{nac}) and propagation along the $x$-axis, yields
\begin{equation}
\frac{\Delta_{d}}{l}=\frac{\mu a_{1} \, c_{2}}{2} \; .
\label{dicro}
\end{equation}
In the particular case of ${\bf{a}}=(a_{1}, 0, a_{3} )$, ${\bf{c}} = (0, c_{2}, 0)$, with the wave propagation along the ${\cal Z}$-axis, the diagonalization 
of the permittivity matrix (\ref{eq68-2}) yields three distinct eigenvalues: $\epsilon, \epsilon \pm i a_{1}c_{2} k /(2\omega^{2})$, which is compatible with the behavior of a biaxial crystal \cite{Robert, Amnon}. For the general case of $\sigma^{B}_{ij}=(a_{i}c_{j}+a_{j}c_{i})/2$, the three distinct eigenvalues are: $\epsilon$, and
\begin{equation}
		\epsilon_{\pm} = \epsilon \pm \frac{i}{2\omega^{2}} \sqrt{ \left[ ({\bf{a}}\times {\bf{c}}) \cdot {\bf{k}} \right]^{2}}.
\end{equation}
As already known, such a scenario is of biaxial materials \cite{Amnon, Robert}, belonging to triclinic, monoclinic, and orthorhombic systems.

\section{Energy propagation}
\label{section3}

In this section, we will discuss some aspects regarding the energy velocity and group velocity considering the three scenarios endowed with magnetic conductivity, $\sigma^{B}_{ij}$, already discussed.

\subsection{\label{energy-velocity-isotropic-case}Isotropic magnetic conductivity}
\begin{figure}[h]
\begin{centering}
\includegraphics[scale=0.69]{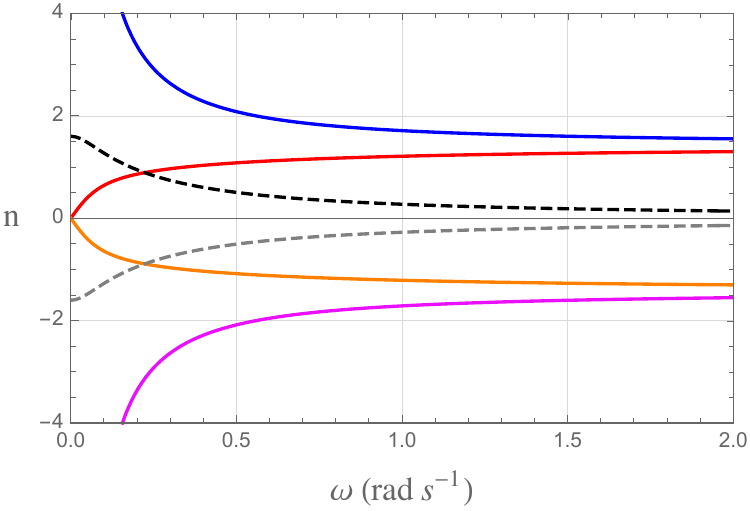}
\par\end{centering}
\caption{Refractive indices of \eqref{isotropic-case-4} in terms of $\omega$. The solid blue (red) line illustrates $\mathrm{Re}[n_{\pm}]$, while the dashed black line depicts $\mathrm{Im}[n_{\pm}]$. The solid orange (magenta) line illustrates $\mathrm{Re}[\tilde{n}_{\pm}]$, while the dashed gray line depicts $\mathrm{Im}[\tilde{n}_{\pm}]$. Here we have used $\mu=1$, $\epsilon'=2$, $\sigma=0.8$ $\mathrm{s}^{-1}$, $\Sigma=0.5$ $\mathrm{s}^{-1}$.}
\label{plot-refractive-indices-isotropic-case}
\end{figure}

Let us begin with the isotropic magnetic conductivity,
$\sigma^{B}_{ij}=\Sigma \, \delta_{ij}$, where $\Sigma$ is a real and positive constant, that
represents $1/3$ of the trace of the $\sigma^{B}_{ij}$ matrix. Substituting this tensor in \eqref{permittivityCME},
the determinant condition (\ref{Det}) yields the $k$-polynomial equation
\begin{align}\label{komegarelationiso}
\left[ \, k^{2} - \mu \, \epsilon(\omega) \, \omega^{2} \, \right]^{2} - \mu^{2} \, \Sigma^{2} \, k^{2}=0 \; .
\end{align}
Remembering that $\epsilon(\omega)=\epsilon^{\prime}(\omega)+i\epsilon^{\prime\prime}(\omega)$, the solutions of \eqref{komegarelationiso} are
\begin{subequations}
\label{isotropic-case-2}
\begin{align}
k_{\pm}&= \omega \left( I_{+} + i \, I_{-} \right) \pm \frac{\mu \Sigma}{2} \; ,
\label{isotropic-case-positive-solutions}
\\
\tilde{k}_{\pm} &= -  \omega \left( I_{+} + i \, I_{-} \right) \pm \frac{\mu \Sigma}{2} \; ,
\label{isotropic-case-negative-solutions}
\end{align}
with
\begin{align}
I_{\pm} &= \frac{\hat{I}_{\pm}}{\sqrt{2}} \sqrt{\mu\epsilon' + \left(\frac{\mu\Sigma}{2\omega}\right)^{2}} \; ,
\label{isotropic-case-3}
\\
\hat{I}_{\pm} &=\sqrt{  \sqrt{ 1 + \left( \frac{\mu \epsilon''}{\mu\epsilon' + \mu^{2}\Sigma^{2} /(4\omega^{2})} \right)^{2}} \pm 1} \; . \label{isotropic-case-3-1}
\end{align}
\end{subequations}
Implementing ${\bf{n}}={\bf{k}}/\omega$ in \eqref{isotropic-case-2}, one obtains the following associated refractive indices:
\begin{subequations}
\label{isotropic-case-4}
\begin{align}
n_{\pm}&=  \left( I_{+} + i \, I_{-} \right) \pm \frac{\mu \Sigma}{2\omega} \; ,
\label{isotropic-case-5} \\
\tilde{n}_{\pm} &= -  \left( I_{+} + i \, I_{-} \right) \pm \frac{\mu \Sigma}{2\omega} \; ,
\label{isotropic-case-6}
\end{align}
\end{subequations}
where $\tilde{n}_{\pm}$ is associated with negative refraction. The behavior of $n_{\pm}$ as a function of the frequency $\omega$ is plotted in Fig. \ref{plot-refractive-indices-isotropic-case}, where for illustration,
we have considered $\epsilon''= i (\sigma/\omega)$, with $\sigma$ being the Ohmic conductivity. Figure \ref{plot-refractive-indices-isotropic-case} also reveals a kind of mirror symmetry between the refractive indices $\tilde{n}_{\pm}$ and $n_{\pm}$, ascribed to $\tilde{n}_{\pm}=-n_{\mp}$. In the following, we use natural units\footnote{ In this paper, we consider natural units. Let us see an example of how to convert from SI units to natural units.
The electric permittivity, $\epsilon$, is measured in {\it farad per meter}, {\it i.e.},
($\mathrm{F} \cdot \mathrm{m}^{-1}$) in SI units. Thus, in natural units where $[\epsilon]=[\mu]=1$, we have
\begin{align}
[\epsilon] &=1 \rightarrow \mathrm{F} \cdot \mathrm{m}^{-1} =1 \rightarrow \mathrm{F}= \mathrm{m} \;   .
\nonumber
\end{align}
Also, in SI units, the conductivity is measured as
\begin{align}
[\sigma] &= \mathrm{\Omega}^{-1} \cdot \mathrm{m}^{-1} = \mathrm{F} \cdot \mathrm{s}^{-1} \cdot
\mathrm{m}^{-1}. \nonumber
\end{align}
Then, by using the previous expressions, we find, in natural units,
\begin{align}
[\sigma] &= \mathrm{\Omega}^{-1} \cdot \mathrm{m}^{-1} = \mathrm{F} \cdot \mathrm{s}^{-1} \cdot
\mathrm{m}^{-1}   \hspace{0.1cm}  \rightarrow \hspace{0.1cm} [\sigma]_{(n.u.)} =  \mathrm{s}^{-1} \; , \nonumber
\end{align}
where the subscript $(n.u.)$ means ``natural units".
For the magnetic conductivity, it analogously holds
\begin{align}
[\sigma^{B}] = \mathrm{\Omega}^{-1} \cdot \mathrm{s}^{-1} = \mathrm{F} \cdot \mathrm{s}^{-1} \cdot
\mathrm{s}^{-1} \hspace{0.1cm} \rightarrow  \hspace{0.1cm}   [\sigma^{B}]_{(n.u.)} =\mathrm{s}^{-1} \; . \nonumber
\end{align}
}.

%
%%%%%%%%%
%

In the presence of absorption terms, the group velocity is not a real quantity, no longer representing the energy propagation velocity. As illustration, we take the dispersion relations (\ref{isotropic-case-positive-solutions}) and (\ref{isotropic-case-negative-solutions}),  and derive the group velocity, $v_{g} =~(\partial \omega / \partial k)$, associated with them, obtaining the expressions
\begin{align}
v_{g}^{\pm}&=\frac{ \displaystyle  \frac{k'_{\pm}}{\mu\omega} \mp \frac{\Sigma}{2\omega} + i \frac{ \omega \epsilon''}{2 k'_{\pm}} \left(1 \mp \frac{\mu \Sigma}{2k'_{\pm}} \right)^{-1}} {\displaystyle \epsilon' + \frac{\omega}{2} \frac{\partial \epsilon'}{\partial \omega} + i \left(\epsilon''+\frac{\omega}{2}\frac{\partial \epsilon''}{\partial \omega} \right)} ,
\label{group-velocity-isotropic-case-1}
\end{align}
\begin{align}
\tilde{v}_{g}^{\pm}&=\frac{ \displaystyle  \frac{\tilde{k}'_{\pm}}{\mu\omega} \mp \frac{\Sigma}{2\omega} + i \frac{ \omega \epsilon''}{2 \tilde{k}'_{\pm}} \left(1 \mp \frac{\mu \Sigma}{2\tilde{k}'_{\pm}} \right)^{-1}} {\displaystyle \epsilon' + \frac{\omega}{2} \frac{\partial \epsilon'}{\partial \omega} + i \left(\epsilon''+\frac{\omega}{2}\frac{\partial \epsilon''}{\partial \omega} \right)} ,
\label{group-velocity-isotropic-case-negative-refraction}
\end{align}
which are complex, as it usually occurs in an absorbing and active media \cite{Gerasik}. In this case, to analyze the propagation of energy carried by the wave through the medium, we evaluate the energy velocity instead of the group velocity. Such velocity is obtained using \eqref{isotropic-case-positive-solutions} and \eqref{isotropic-case-negative-solutions} in \eqref{vE}, for the propagating modes described by $k_{\pm}$ and $\tilde{k}_{\pm}$, namely:
\begin{align}
V_{E}^{\pm} &=  \frac{ 2\omega I_{+} \pm {\mu \Sigma} }{ 2\omega U^{I}_{\pm} } \; ,
\quad
\widetilde{V}_{E}^{\pm} =  \frac{- 2\omega I_{+} \pm \mu \Sigma} { 2\omega U^{I}_{\mp} } \; ,
\label{energy-velocity-isotropic-case-1}
\end{align}
with
\begin{align}
U^{I}_{\pm} &=  \frac{\mu \epsilon'}{2} + \frac{\mu\omega}{2}\frac{ \partial \epsilon'}{\partial \omega} + \frac{1}{2} (I_{+}^{2} + I_{-}^{2} ) + \frac{ \mu^{2} \Sigma^{2}}{8\omega^{2}} \pm \frac{ \mu \Sigma I_{+}}{2\omega}. \label{energy-velocity-isotropic-case-extra-1}
\end{align}

\begin{figure}[h]
\begin{centering}
\includegraphics[scale=0.69]{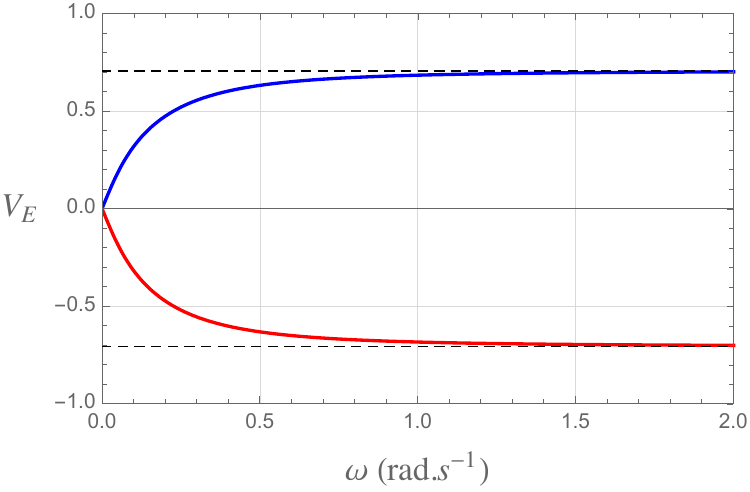}
\par\end{centering}
\caption{\label{plot-energy-velocity-isotropic-case} Energy velocity in terms of $\omega$. The solid blue (red) line illustrates $V_{E}^{\pm}$ and $\widetilde{V}_{E}^{\pm}$, while the horizontal dashed black lines depict the asymptotic limits $\pm1/\sqrt{\mu\epsilon'}$. Here we have used $\mu=1$, $\epsilon'=2$, $\sigma=0.8$ $\mathrm{s}^{-1}$, $\Sigma=0.5$ $\mathrm{s}^{-1}$.}
\end{figure}

Clearly, the energy velocity (\ref{energy-velocity-isotropic-case-1}) is different from the group velocity (\ref{group-velocity-isotropic-case-1}), $V_{E}^{\pm} \neq v_{g}^{\pm}$. Such an energy velocity is plotted as a function of the frequency in Fig.~\ref{plot-energy-velocity-isotropic-case}, with the values of $\mu=1$, $\epsilon'=2$, $\sigma=0.8$ $s^{-1}$, and $\Sigma=0.5$ $s^{-1}$. For the high-frequency limit, $\omega \rightarrow \infty$, the energy velocity goes as $v_{E}= \pm \, 1/\sqrt{\mu\epsilon}$, with the plus sign holding for the propagating modes associated with $k_{\pm}$, and the minus sign being related to $\tilde{k}_{\pm}$.  Notice that when there is no magnetic conductivity ($\Sigma=0)$, the group velocity (\ref{group-velocity-isotropic-case-1}) recovers the same result obtained in Ref.~\cite{Gerasik}.
For non-absorbing dielectrics ($\epsilon''=0$), the group and energy velocities are usually equal, $v_{E} = v_{g}$
\cite{Gerasik,Loudon,Zangwill, Ruppin}. Such equality fails, however, in the presence of magnetic conductivity, $\Sigma \neq 0$. Indeed, for $\epsilon''=0$, one finds $I_{-}=k_{\pm}''=0$, $I_{+}= \sqrt{\mu\epsilon' + \mu^{2}\Sigma^{2}/(4\omega^{2})}$, and $k'_{\pm}=\omega I_{+} \pm \mu \Sigma/2$, in which the group velocity (\ref{group-velocity-isotropic-case-1}) and the energy velocity (\ref{energy-velocity-isotropic-case-1}), respectively, read

\begin{subequations}
\begin{align}
	 \left. v_{g}^{\pm} \right|_{\epsilon^{\prime\prime}= 0}&=  \frac{2 k'_{\pm} \mp \mu \Sigma}
	{\displaystyle 2\mu \omega \left( \epsilon' + \frac{\omega}{2} \frac{\partial \epsilon'}{\partial \omega} \right) }   \; ,
	\label{isotropic-case-8}
	\\
\left. V_{E}^{\pm} \right|_{\epsilon^{\prime\prime}= 0}&= \frac{ k'_{\pm}}{ \displaystyle \mu\omega \left( \epsilon' + \frac{\omega}{2} \frac{\partial \epsilon'}{\partial \omega} \pm \frac{\Sigma}{2\omega^{2}} k_{\pm}' \right)} \; ,
\label{isotropic-case-7}
\end{align}
\end{subequations}
constituting distinct expressions, $v_{E} \neq v_{g}$, even when the dielectric substrate is deprived of absorption $(\epsilon''=0)$. This is an entire consequence of the isotropic magnetic conductivity, which can not be encoded in the medium permittivity. In fact, note that by setting $\Sigma =0$, \eqref{isotropic-case-8} and \eqref{isotropic-case-7} reduce to the same expression,
\begin{align}\label{isotropic-case-Sigma0}
V_{E}= V_{g}=\frac{ k'_{\pm}}{ \mu\omega \left( \epsilon' + \frac{1}{2} \omega ({\partial \epsilon' /\partial \omega}) \right)}  \; .
\end{align}

This is a curious result in some respects. In fact, on the one hand, it seems to indicate that the magnetic conductivity implies absorption. On the other hand, equations (\ref{n23-1A}) and (\ref{n23-1B}) reveal that the $\Sigma$ conductivity does not engender an imaginary contribution for the refractive indices, holding the absence of absorption (for $\epsilon''=0$ and $\Sigma \neq0$). Therefore, it here appears a novelty: a non-absorptive medium (real refractive index) where the energy velocity and group velocity are unequal. This surprising behavior is due to the presence of the isotropic magnetic current, which renders these velocities extremely discrepant at low frequencies, as seen below.

Concerning the negative refraction modes associated with $\tilde{k}_{\pm}$, given in Eq. (\ref{isotropic-case-negative-solutions}), for $\epsilon''=0$, one writes the velocities,
\begin{subequations}
	\begin{align}
		\left. \tilde{v}_{g}^{\pm} \right|_{\epsilon^{\prime\prime}= 0}&=  \frac{2 \tilde{k}'_{\pm} \mp \mu \Sigma}
		{\displaystyle 2\mu \omega \left( \epsilon' + \frac{\omega}{2} \frac{\partial \epsilon'}{\partial \omega} \right) }   \; ,
		\label{isotropic-case-8B}
		\\
		\left. \tilde{V}_{E}^{\pm} \right|_{\epsilon^{\prime\prime}= 0}&= \frac{ \tilde{k}'_{\pm}}{ \displaystyle \mu\omega \left( \epsilon' + \frac{\omega}{2} \frac{\partial \epsilon'}{\partial \omega} \pm \frac{\Sigma}{2\omega^{2}} \tilde{k}_{\pm}' \right)} \; .
		\label{isotropic-case-7B}
	\end{align}
\end{subequations}
For the group velocities (\ref{isotropic-case-8}) and (\ref{isotropic-case-8B}), it holds that $v_{g}^{+}=v_{g}^{-}$ and $\tilde{v}_{g}^{+}=\tilde{v}_{g}^{-}$. This happens due to the presence of the factor  $\mp \mu \Sigma /2$ in the numerator of these relations and the factor $\pm \mu \Sigma /2$ contained inside $k'_{\pm}$ and $\tilde{k}'_{\pm}$ [vide Eqs.~(\ref{isotropic-case-positive-solutions}) and (\ref{isotropic-case-negative-solutions}), respectively], yielding a cancelation, so that the group velocities are
\begin{subequations}
\label{group-velocities-isotropic-case-1}
\begin{align}	
\label{isotropic-case-8-1}
	\left. v_{g}^{\pm} \right|_{\epsilon^{\prime\prime}= 0}&=  \frac{  \sqrt{\mu
			\epsilon' + \mu^{2}\Sigma^{2} /(4 \omega^{2})}}{\displaystyle \mu \epsilon' + \frac{\mu\omega}{2} \frac{\partial \epsilon'}{\partial \omega}  }, \\
	\left. \tilde{v}_{g}^{\pm} \right|_{\epsilon^{\prime\prime}= 0}&= - \frac{  \sqrt{\mu
			\epsilon' + \mu^{2}\Sigma^{2} /(4 \omega^{2})}}{\displaystyle \mu \epsilon' + \frac{\mu\omega}{2} \frac{\partial \epsilon'}{\partial \omega}  }.
	\label{isotropic-case-8-2}
\end{align}
\end{subequations}
Moreover, from \eqref{isotropic-case-7} and \eqref{isotropic-case-7B} one also finds,
\begin{subequations}
	\label{isotropic-case-8-3}
	\begin{align}
		V_{E}^{+}-V_{E}^{-}= \frac{2\omega^{2} \Sigma}{F} \frac{\partial \epsilon'}{\partial \omega}, \quad 	\tilde{V}_{E}^{+}-\tilde{V}_{E}^{-}= \frac{2\omega^{2} \Sigma}{F} \frac{\partial \epsilon'}{\partial \omega}, \label{isotropic-case-8-4A}
	\end{align}
	where
	\begin{align}
		F&={ 4 \epsilon^{2} \omega^{2} + \mu \epsilon \Sigma^{2} + \omega^{4} \left( \frac{\partial \epsilon'}{\partial \omega} \right)^{2} + \omega \frac{ \partial \epsilon'}{\partial \omega}  \left( \mu \Sigma^{2}+4 \epsilon \omega^{2} \right) }  . \label{isotropic-case-8-4B}
	\end{align}
\end{subequations}
Hence, with $\epsilon'$ constant, it holds $V_{E}^{+}=V_{E}^{-}$ and $\tilde{V}_{E}^{+}=\tilde{V}_{E}^{-}$.

\begin{figure}[H]
\begin{centering}
\includegraphics[scale=0.68]{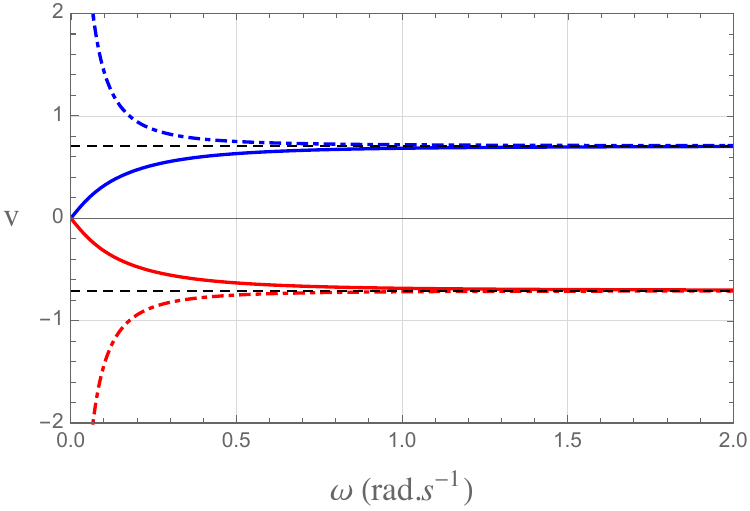}
\par\end{centering}
\caption{\label{comparison-group-energy-velocities-isotropic-case} Group velocities of \eqref{isotropic-case-8-1} and \eqref{isotropic-case-8-2}. Energy velocities of \eqref{isotropic-case-7} and \eqref{isotropic-case-7B}. Solid lines indicates $v_{E}^{\pm}$ (blue) and $\tilde{v}_{E}^{\pm}$ (red). Dashed-dotted curves represent  $v_{g}^{\pm}$ (blue) and $\tilde{v}_{g}^{\pm}$ (red). Here, we have used $\mu=1$, $\epsilon'=2$, $\epsilon''=0$, $\Sigma=0.5$ $\mathrm{s}^{-1}$.}
\end{figure}

The behavior of the velocities $v_{g}^{\pm}$, $V_{E}^{\pm}$,  $\tilde{v}_{g}^{\pm}$ and $\tilde{V}_{E}^{\pm}$, for $\epsilon'=cte$ and $\epsilon''=0$, is depicted in Figure \ref{comparison-group-energy-velocities-isotropic-case}. In such a plot, one notices the extreme distinction between group velocity $v_{g}$ and energy velocity $V_{E}$ at low frequencies. The group velocity tends to infinity at the origin, being superluminal at a certain frequency range. At large frequencies, the group velocity approaches the energy velocity. Despite its real character, the behavior of the group velocity for low frequencies is an illustration of its limitations in representing the propagation of energy or signal in scenarios with strong anomalous dispersion.

\subsection{\label{antisymmetric-case-section}Antisymmetric magnetic conductivity}
\begin{figure}[H]
\begin{centering}
\includegraphics[scale=0.69]{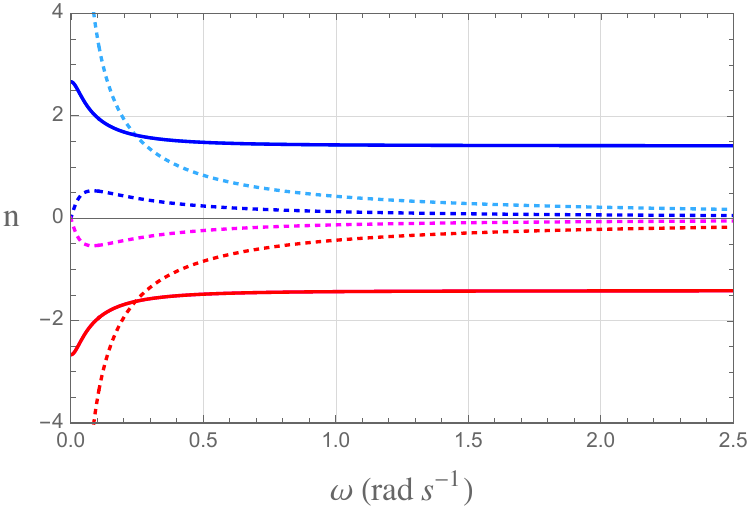}
\par\end{centering}
\caption{\label{plot-refractive-indices-antisymmetric-case} Refractive indices $n_{\pm}$ of \eqref{antisymmetric-case-5}. Solid (dotted) lines represent $\mathrm{Re}[n_{\pm}]$ ($\mathrm{Im}[n_{\pm}]$). For $n_{+}$, we have considered $\theta=0$ (blue) and $\theta=\pi$ (cyan), and for $n_{-}$, $\theta=0$ (red) and $\theta=\pi$ (magenta). Here we have used $\mu=1$, $\epsilon'=2$, $\sigma=0.8$~$\mathrm{s}^{-1}$, $b=0.3$ $\mathrm{s}^{-1}$. The solid blue and cyan lines lie on top of each other, the same occurring for the solid red and magenta lines.}
\end{figure}

In this case, the antisymmetric conductivity tensor is parametrized in terms of a
constant background vector, ${\bf b}$, as previously mentioned, that is, $\sigma_{ij}^{B}=\epsilon_{ijk} \, b_{k}$.
%
%\begin{align}
%\label{sigmaBsymm}
%\sigma_{ij}^{B}=\epsilon_{ijk} \, b_{k} \; .
%\end{align}
%
Substituting this antisymmetric tensor in (\ref{permittivityCME}), the determinant (\ref{Det}) yields the dispersion relation
\begin{align}
\label{realdispantysim}
k^{2} + i \, \mu \, ({\bf b}\cdot {\bf k}) - \mu \, \epsilon(\omega) \, \omega^{2} = 0 \; ,
\end{align}
whose solutions are
\begin{align}\label{kantisymmetric}
k_{\pm} = -\frac{i\,\mu}{2} \, (\hat{{\bf k}}\cdot {\bf b})
\pm \sqrt{\mu\,\epsilon(\omega)\,\omega^2-\frac{\mu^2}{4}(\hat{{\bf k}}\cdot {\bf b})^2 }
\; ,
\end{align}
or equivalently,
\begin{subequations}
\label{antisymmetric-case-1}
\begin{align}
k_{\pm} &= - \frac{i\,\mu}{2} \, (\hat{\bf{k}}\cdot {\bf{b}}) \pm \omega  \left(A_{+} + i \, A_{-}\right) \; ,
\label{antisymmetric-case-2}
\end{align}
with
\begin{align}
A_{\pm} &= \frac{1}{\sqrt{2}} \, \sqrt{ \left| f(\omega) \right| \left(\sqrt{ 1 +  \frac{\mu^{2}  \epsilon''^{2}}{{f(\omega)}^2}}
\pm \mathrm{sign}[f(\omega)]\right) } \;  ,
\label{antisymmetric-case-3}
\end{align}
where
\begin{align}
f(\omega)&=\mu \epsilon' - \frac{\mu^{2} (\hat{\bf{k}}\cdot {\bf{b}})^{2}}{4\omega^{2}} \; .
\label{antisymmetric-case-4}
\end{align}
\end{subequations}
The refractive indices, $n_{\pm}$, associated with $k_{\pm}$ can be obtained by using ${\bf{n}}={\bf{k}}/\omega$, providing two indices
\begin{align}
n_{\pm} &= -i \, \frac{\mu (\hat{\bf{k}}\cdot {\bf{b}})}{2\omega} \pm (A_{+} + i A_{-} ) \; ,
\label{antisymmetric-case-5}
\end{align}
being $\mathrm{Re}[n_{+}]>0$ and $\mathrm{Re}[n_{-}]<0$ (negative refraction).
Using $(\hat{\bf{k}}\cdot {\bf{b}})=b\,\cos\theta$ and $\epsilon(\omega)=\epsilon'+ i (\sigma/\omega)$, we illustrate the general behavior of the refractive index $n_{\pm}$ in terms of the frequency in Fig.~\ref{plot-refractive-indices-antisymmetric-case} for the following cases: i) parallel ($\theta=0$) and ii) antiparallel ($\theta=\pi$) configurations. We use $\mu=1$, $\epsilon=2$, $|{\bf b}|=0.3$ $\mathrm{s}^{-1}$, and $\sigma=0.8$ $\mathrm{s}^{-1}$ in these plots. Notice that, when ${\bf b}$ is perpendicular to $\hat{{\bf k}}$-direction, the results do not depend on the magnitude of the ${\bf b}$-vector.

We point out that $\mathrm{Re}[n_{\pm}]|_{\theta=0} = \mathrm{Re}[n_{\pm}]|_{\theta=\pi}$, explaining the reason by which solid blue and cyan curves for $\mathrm{Re}[n_{+}]$, and magenta and red curves for $\mathrm{Re}[n_{-}]$ appear superimposed in Fig.~\ref{plot-refractive-indices-antisymmetric-case}. Furthermore, one notices that $\mathrm{Im}[n_{-}]|_{\theta=0}=-\mathrm{Im}[n_{+}]|_{\theta=\pi}$ and $\mathrm{Im}[n_{-}]|_{\theta=\pi}=-\mathrm{Im}[n_{+}]|_{\theta=0}$, which indicates an interchange in the absorptive behavior for $\bf{b}$ parallel and antiparallel configurations. The correspondence $\mathrm{Re}[n_{\pm}] = -\mathrm{Re}[n_{\mp}]$ holds for any propagation configuration. In the high-frequency limit,  $n_{\pm} \rightarrow\pm \sqrt{\mu\epsilon'}$, which also indicates that the absorption effects reduce drastically in this special limit.
To evaluate the energy velocity associated with $k_{\pm}$, one uses the dispersion relations (\ref{antisymmetric-case-2}) in \eqref{vE}, yielding
\begin{align}
V_{E}^{\pm} &= \frac{  \pm A_{+}} { U^{A}_{\pm}} \; , \label{antisymmetric-case-7}
\end{align}
with
\begin{align}
U^{A}_{\pm} &=\frac{\mu\epsilon'}{2} + \frac{\mu\omega}{2} \frac{\partial \epsilon'}{\partial \omega} + \frac{1}{2} (A_{+}^{2} + A_{-}^{2}) + \frac{\mu^{2} (\hat{\bf{k}}\cdot {\bf{b}})^{2}}{8\omega^{2}} + \nonumber \\
&\phantom{=}\mp \frac{\mu (\hat{\bf{k}}\cdot {\bf{b}}) A_{-}}{2\omega} \; .
\label{antisymmetric-case-7-1}
\end{align}

effectively

\begin{figure}[h]
\begin{centering}
\includegraphics[scale=0.69]{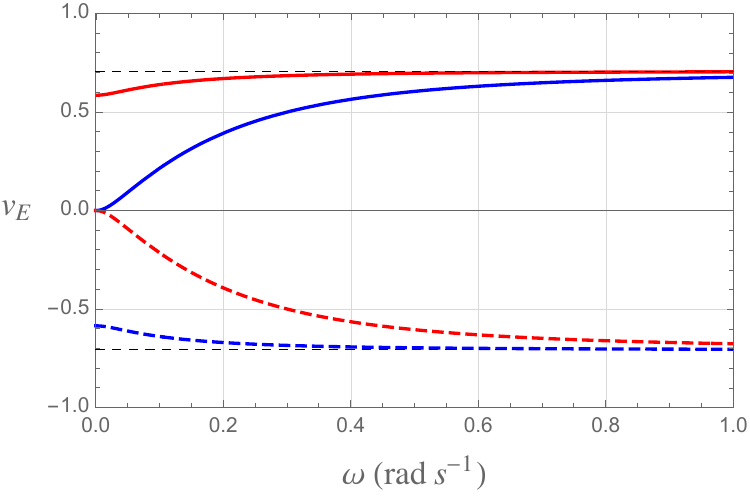}
\par\end{centering}
\caption{\label{plot-energy-velocities-antisymmetric-case-version-2} Energy velocity $V_{E}^{\pm}$ of \eqref{antisymmetric-case-7}. The particular cases considered are: $\theta=0$ (red) and $\theta=\pi$ (blue). Solid lines indicate $V_{E}^{+}$, while the dashed curves represent $V_{E}^{-}$. Here we have used $\mu=1$, $\epsilon'=2$, $\sigma=0.8$~$\mathrm{s}^{-1}$, $b_{3}=0.3$ $\mathrm{s}^{-1}$.}
\end{figure}

We show the behavior of the energy velocity $V_{E}^{\pm}$ as a function of the frequency in Fig.~\ref{plot-energy-velocities-antisymmetric-case-version-2}, for two cases, $\theta=0$ and $\theta=\pi$, which fulfill the mode orthogonality conditions given in Eqs. (\ref{complex-group-velocity-isotropic-54A}) and (\ref{complex-group-velocity-isotropic-54B}).
Taking the differential operation in relation to $k^{i}$-component in (\ref{realdispantysim}), the correspondent group velocity is
\begin{align}\label{vgaintsymmetick}
{\bf v}_{g} = \frac{\partial \omega}{\partial {\bf k}} = \frac{ \displaystyle  \frac{{\bf k}}{\mu \omega} + i \, \frac{{\bf b}}{2\omega}}{ \displaystyle \left(\epsilon' + \frac{\omega}{2} \frac{ \partial \epsilon'}{\partial \omega} \right) + i \left(\epsilon'' + \frac{\omega}{2} \frac{ \partial \epsilon''}{\partial \omega} \right)} \; ,
\end{align}
which is valid for all frequency ranges and for both solutions, $k_{\pm}$. Let us now write the $k^{i}$-components as $k^{i} = k'^{i} + i \, k''^{i}$, so that the group velocity (after an algebraic development) can be rewritten as
\begin{align}
\label{vgantisimetric}
{\bf v}_{g} &= \frac{ \displaystyle \frac{{\bf k}^{\prime}}{\mu\omega} + i \, \frac{\omega \epsilon'' \, \hat{{\bf k}} }{2 k'}} { \displaystyle \left( \epsilon' + \frac{\omega}{2} \frac{\partial \epsilon'}{\partial \omega} \right) + i \left(\epsilon'' + \frac{\omega}{2} \frac{\partial \epsilon''}{\partial \omega} \right) }  \; .
\end{align}
Therefore, the group velocity is in the ${\bf k}$-direction since the contribution
of the ${\bf b}$-vector conductivity is canceled.
Differently from the isotropic conductivity case, where the absorption effect occurs only for $\epsilon''\neq 0$, in the antisymmetric scenario,
the absorption effect can be observed by means of the dispersion relation (\ref{realdispantysim}), from which one finds that the absorption occurs
for the three subcases:
\begin{itemize}
 \item[i)]  $\epsilon'' =0$ and $b\cos\theta=0$ (absence of absorption);
 \item[ii)] $\epsilon''=0$ and $b\cos\theta \neq 0$ (\textit{magnetically induced absorption effect});
  \item[iii)] $\epsilon''\neq 0$ and $b\cos\theta \neq 0$ (usual and magnetically induced absorption), for which group and energy velocities are different, as revealed by Eqs. (\ref{antisymmetric-case-7}) and (\ref{vgantisimetric}).
\end{itemize}

 In order to analyze the energy velocity and its possible connection with the group velocity, we now consider the two particular cases with potential novelties, that is, the ones with no dielectric absorption,  $\epsilon'' =0$.
For the case (i), $\epsilon''=0$ and $b\cos\theta=0$, the complex group velocity of (\ref{vgantisimetric}) and the energy velocity (\ref{antisymmetric-case-7}) are equivalent:
\begin{align}
\label{vg=vE}
v_{g}^{\pm}=V_{E}^{\pm}= \frac{ {k'_{\pm} /\mu\omega}}{\displaystyle  \epsilon' + \frac{\omega}{2}\frac{\partial \epsilon'}{\partial \omega} } \; ,
\end{align}
with
\begin{equation}
k'_{\pm}= \omega A_{+}, \quad k''_{\pm}=0, \quad  A_{+}=\sqrt{\mu\epsilon'}, \quad A_{-}=0 \; .
\end{equation}
This is an expected result, considering the real refractive index, $n=\sqrt{\mu\epsilon'}$, that stems from \eqref{eq68}.
For the case (ii), $\epsilon''=0$ and $b\cos\theta \neq0$, the group velocity becomes real and equal to the energy velocity, as below
\begin{align}\label{vgvE}
v_{g}^{\pm}=V_{E}^{\pm}= \frac{{k'_{\pm} /\mu\omega}}{ \displaystyle \epsilon' + \frac{\omega}{2}\frac{\partial \epsilon'}{\partial \omega} } \; ,
\end{align}
where, in this case,
\begin{subequations}
\label{particular-case-antisymmetric-extra-0}
\begin{align}
k'_{\pm}&= \omega A_{+}, \quad k''_{\pm}=-\frac{\mu}{2} (\hat{\bf{k}}\cdot {\bf{b}}) \pm \omega A_{-} \; ,
\label{particular-case-antisymmetric-extra-1}
\end{align}
\begin{align}
A_{\pm} =\frac{1}{\sqrt{2}}\sqrt{ \left| f(\omega) \right| \left(1 \pm \mathrm{sgn}[f(\omega)]\right) } \; ,
\label{antisymmetric-case-3b}
\end{align}
\end{subequations}
so that
\begin{align}	\label{vg=vE}
v_{g}^{\pm}=V_{E}^{\pm}= \frac{ \sqrt{ \left| f(\omega) \right| \left(1 + \mathrm{sgn}[f(\omega)]\right) }}{ \displaystyle \mu \sqrt{2} \left( \epsilon' + \frac{\omega}{2}\frac{\partial \epsilon'}{\partial \omega} \right)} \; ,
\end{align}
with $f(\omega)$ given by \eqref{antisymmetric-case-4}. Differently from the isotropic conductivity section, the group velocity and energy velocity remain equal even when the anisotropic conductivity is non-null. It is interesting to observe that when $\epsilon''=0$ and $b\cos\theta \neq 0$, there is absorption, since the refractive index (\ref{eq68}),
\begin{equation}\label{eq68b}
n_{\pm}=\pm\sqrt{\mu\epsilon^{\prime}-\left( \frac{\mu}{2\omega}b\cos\theta  \right)^2}+\mathrm{i}\frac{\mu}{2\omega}b\cos\theta \, ,
\end{equation}
 possesses an imaginary piece. Surprisingly, however, the group velocity and energy velocity turn out equivalent in this special absorbing scenario. Therefore, the equality $v_{g}=V_{E}$ holds for the magnetically induced absorption, stemming from the antisymmetric magnetic conductivity tensor.

\begin{figure}[h]
\begin{centering}
\includegraphics[scale=0.69]{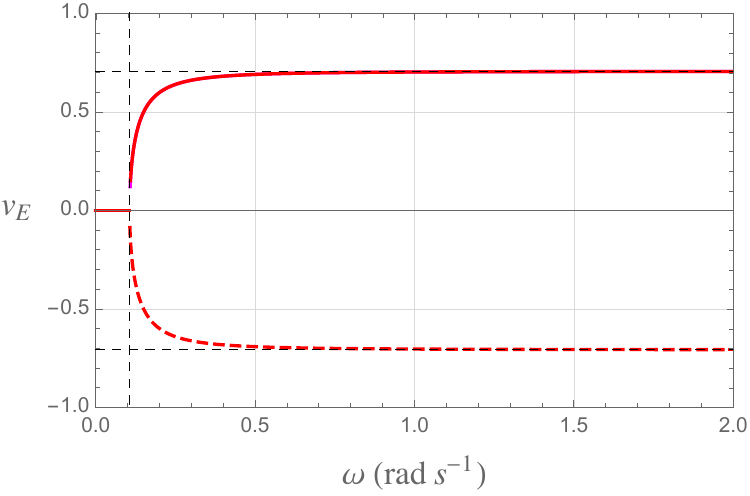}
\par\end{centering}
\caption{\label{plot-energy-velocity-antisymmetric-case-null-imaginary-part-of-epsilon} Energy velocity $V_{E}^{\pm}$ of \eqref{vg=vE} for $\epsilon''=0$. Solid red line indicates $V_{E}^{+}$ for $\theta=(0,\pi)$, while the dashed red curve represents $V_{E}^{-}$ for $\theta=(0,\pi)$. Here we have used $\mu=1$, $\epsilon'=2$, $b=0.3$ $\mathrm{s}^{-1}$. The vertical dashed line indicates the frequency $\omega'$ of \eqref{cutoff-frequency-antisymmetric-case}.}
\end{figure}

Here, we need to be careful due to the sign function in \eqref{vg=vE}, as highlighted below:
\begin{itemize}
\item For $f(\omega) >0$, one has $\mathrm{sign}\left[f(\omega) \right]=1$, such that
\begin{align}
v_{g}^{\pm}=V_{E}^{\pm}= \frac{ \sqrt{ \left| f(\omega) \right| }}{ \displaystyle \mu \left( \epsilon' + \frac{\omega}{2}\frac{\partial \epsilon'}{\partial \omega} \right)} \; .
\label{extra-observation-1}
\end{align}
\item For $f(\omega) <0$, one has $\mathrm{sgn}\left[f(\omega) \right]=-1$, then one finds
\begin{align}
v_{g}^{\pm}=V_{E}^{\pm}= 0 \; ,
\label{extra-observation-2}
\end{align}
which is a consequence of having the wave vector (\ref{kantisymmetric}) completely imaginary (for this condition).
\end{itemize}
Such behavior is plotted in Fig. \ref{plot-energy-velocity-antisymmetric-case-null-imaginary-part-of-epsilon}, that depicts the energy velocity $V_{E}^{\pm}$ considering $\epsilon''=0$ for the parallel and antiparallel configurations. For $0<\omega<\omega'$, with
\begin{align}
\omega'=\sqrt{ \frac{\mu (\hat{\bf{k}}\cdot {\bf{b}})^{2}}{4 \epsilon'}} \; ,
\label{cutoff-frequency-antisymmetric-case}
\end{align}
it holds $f(\omega) <0$ and energy velocity is null, indicating that, in this particular situation, there is no propagation. For $\omega>\omega'$, one has $f(\omega) >0$ and the velocities (\ref{extra-observation-1}).

\subsection{\label{symmetric-case-section}Symmetric magnetic conductivity}
The magnetic conductivity tensor in the symmetric form is parameterized by
$\sigma^{B}_{ij}=\left( a_{i} \, c_{j} + a_{j} \, c_{i} \right)/2$,
%
%\begin{align}\label{sigmaBsymm}
%\sigma^{B}_{ij}=\frac{1}{2} \left( a_{i} \, c_{j} + a_{j} \, c_{i} \right) \; ,
%\end{align}
%
where $a_{i}$ and $c_{j}$ are components of the two background vectors ${\bf a}$ and ${\bf c}$, respectively.
In this case, the permittivity tensor (\ref{permittivityCME}) is
\begin{align}\label{barepsilonijsymm}
\overline{\epsilon}_{ij}(\omega)= \epsilon(\omega) \, \delta_{ij}
+\frac{i}{2\omega} \left[ \, a_{i} \left( {\bf c} \times {\bf n} \right)_{j}
+ c_{i} \left( {\bf a} \times {\bf n} \right)_{j} \, \right] \; .
\end{align}
The null determinant in (\ref{Det}) yields the $k$-polynomial equation
\begin{align}
k^2  \pm  \frac{i\,\mu}{2}  \, ({\bf a}\times{\bf c}) \cdot {\bf k} \, - \, \mu \,\epsilon(\omega)\,\omega^2
&= 0 \; ,
\label{Eqsnsymm1}
\end{align}
whose four solutions are
\begin{subequations}
\label{symmetric-case-1}
\begin{align}
k_{\pm} &= \pm \frac{i \mu ({\bf{a}} \times {\bf{c}})\cdot\hat{\bf{k}}}{4}  + \omega (S_{+} + i S_{-}) \; ,
\label{symmetric-case-2}
\\
\tilde{k}_{\pm}&= \pm \frac{i \mu ({\bf{a}} \times {\bf{c}})\cdot\hat{\bf{k}}}{4} - \omega (S_{+} + i S_{-}) \; ,
\label{symmetric-case-3}
\end{align}
with
\begin{align}
S_{\pm} &= \frac{1}{\sqrt{2}} \sqrt{ |g(\omega)| \left( \sqrt{1 + \frac{\mu^{2}\epsilon''^{2}}{g(\omega)^{2}}} \pm \mathrm{sgn}[g(\omega)] \right) } \; , \label{symmetric-case-4} \\
g(\omega) &= \mu\epsilon'-\frac{\mu^{2}}{16\omega^{2}} (({\bf{a}} \times {\bf{c}})\cdot\hat{\bf{k}})^{2} \; .
\label{symmetric-case-5}
\end{align}
\end{subequations}
Using $({\bf{a}}\times {\bf{c}} ) \cdot \hat{\bf{k}} = |{\bf{a}}\times {\bf{c}}| \cos\varphi$, there appear four refractive indices,
\begin{subequations}
\label{refractive-indices-symmetric-case-6-0}
\begin{align}
n_{\pm} &=\pm \frac{i \mu |{\bf{a}} \times {\bf{c}}|}{4\omega} \cos\varphi + (S_{+} + i S_{-}) \; ,
\label{symmetric-case-6}
\\
\tilde{n}_{\pm}&=\pm \frac{i \mu |{\bf{a}} \times {\bf{c}}|}{4\omega} \cos\varphi - (S_{+} + i S_{-}) \; ,
\label{symmetric-case-7}
\end{align}
\end{subequations}
where $\tilde{n}_{\pm}$ are related with negative refraction.

\begin{figure}[H]
\begin{centering}
\includegraphics[scale=0.69]{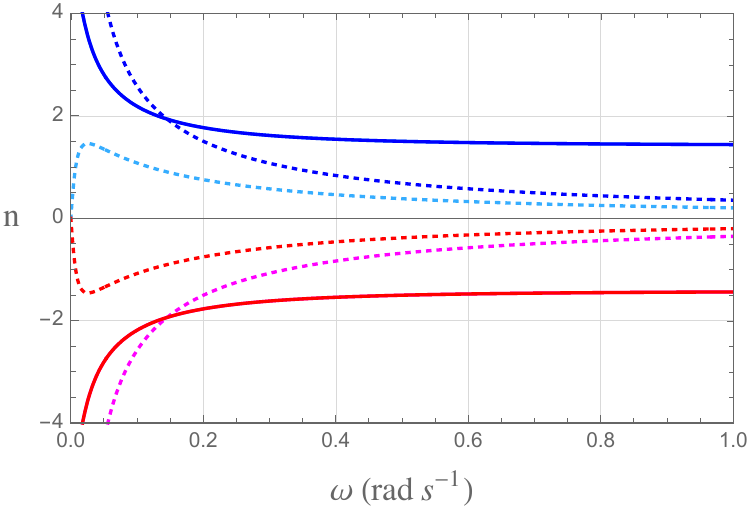}
\par\end{centering}
\caption{\label{plot-refractive-indices-symmetric-case} Refractive indice $n_{+}$ of \eqref{symmetric-case-6} and $\tilde{n}_{+}$ of \eqref{symmetric-case-7}. Solid (dashed) lines represent $\mathrm{Re}[n_{+}, \tilde{n}_{+}]$ ($\mathrm{Im}[n_{+}, \tilde{n}_{+}]$). For $n_{+}$, we have considered $\theta=0$ (blue) and $\theta=\pi$ (cyan), and for $\tilde{n}_{+}$, $\theta=0$ (red) and $\theta=\pi$ (magenta).  Here, we have used $\mu=1$, $\epsilon'=2$, $\sigma=0.8$ $\mathrm{s}^{-1}$, $|{\bf{a}}\times {\bf{c}}|=0.3$~$\mathrm{s}^{-1}$. The solid red and magenta lines lie on top of each other, and solid blue and cyan curves are coincident, since $\cos^2 \pi=\cos^2 0.$ }
\end{figure}

Figure \ref{plot-refractive-indices-symmetric-case} illustrates the general behavior of the refractive indices $n_{\pm}$ and $\tilde{n}_{\pm}$ as functions of the frequency for: parallel ($\varphi=0$) and antiparallel ($\varphi=\pi$) configurations, which are the ones compatible with the orthogonality conditions (\ref{complex-group-velocity-isotropic-54A}) and (\ref{complex-group-velocity-isotropic-54B}).

\begin{figure}[H]
\begin{centering}
\includegraphics[scale=0.69]{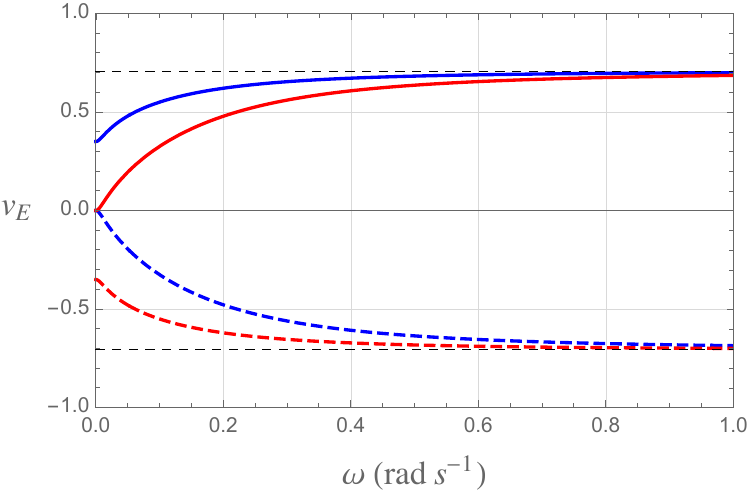}
\par\end{centering}
\caption{\label{plot-energy-velocity-symmetric-case-vplus-vtilplus} Energy velocity $V_{E}^{+}$ and $\widetilde{V}_{E}^{+}$ of \eqref{symmetric-case-12}. The particular cases considered here are: $\theta=0$ (red) and $\theta=\pi$ (blue). Solid (dashed) lines indicate $V_{E}^{+} (\widetilde{V}_{E}^{+})$, respectively. Here, we have used $\mu=1$, $\epsilon'=2$, $\sigma=0.8$~$\mathrm{s}^{-1}$, $ |{\bf{a}} \times {\bf{c}}|=0.3$~$\mathrm{s}^{-1}$.}
\end{figure}

\begin{figure}[ht]
\begin{centering}
\includegraphics[scale=0.69]{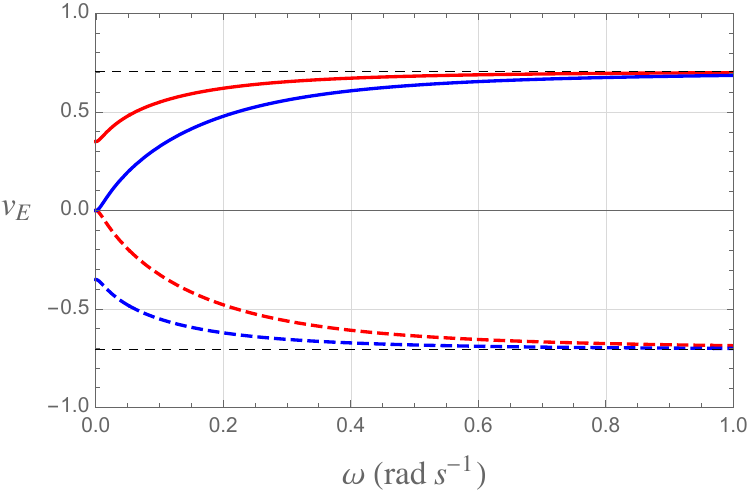}
\par\end{centering}
\caption{\label{plot-energy-velocity-symmetric-case-vminus-vtilminus} Energy velocity $V_{E}^{-}$ and $\widetilde{V}_{E}^{-}$ of \eqref{symmetric-case-12}. The particular cases considered are:  $\theta=0$ (red) and $\theta=\pi$ (blue). Solid (dashed) curves represent $V_{E}^{-}$ $(\widetilde{V}_{E}^{-})$. Here we have used $\mu=1$, $\epsilon'=2$, $\sigma=0.8$~$\mathrm{s}^{-1}$, $ |{\bf{a}} \times {\bf{c}}|=0.3$~$\mathrm{s}^{-1}$. }
\end{figure}

From \eqref{symmetric-case-6}, one observes that $\mathrm{Re}[n_{+}] = \mathrm{Re}[n_{-}]$, for all $\varphi$. Also, $\mathrm{Im}[n_{-}]|_{\varphi=\pi} = \mathrm{Im}[n_{+}]|_{\varphi=0}$ and $\mathrm{Im}[n_{-}]|_{\varphi=0} = \mathrm{Im}[n_{+}]|_{\varphi=\pi}$, which indicates that the absorbing terms in $n_{+}$ for the parallel configuration are mapped in absorbing terms in $n_{-}$ for the antiparallel configuration. This kind of analysis allows us to infer the behavior of $n_{-}$ without plotting it.
Comparing the refractive indices $n_{\pm}$ and $\tilde{n}_{\pm}$, one notices that $\mathrm{Re}[\tilde{n}_{\pm}] = -S_{+} = - \mathrm{Re}[n_{\pm}]$.
On the other hand,
\begin{subequations}
\begin{align}
-\mathrm{Im}[\tilde{n}_{\pm}]&=S_{-}\mp i \frac{\mu}{4\omega} |{\bf{a}}\times {\bf{c}}| \cos\varphi \; ,
\label{symmetric-extra-2} \\
\mathrm{Im}[{n}_{\pm}]&=S_{-}\pm i \frac{\mu}{4\omega} |{\bf{a}}\times {\bf{c}}| \cos\varphi \; ,
\label{symmetric-extra-1}
\end{align}
\end{subequations}
which implies
\begin{align}
\mathrm{Im}[\tilde{n}_{\pm}]|_{\varphi}=-\mathrm{Im}[n_{\pm}]|_{\pi-\varphi} \; .
\label{symmetric-extra-2}
\end{align}
For the solutions $k_{\pm}$ of \eqref{symmetric-case-6} and $\tilde{k}_{\pm}$ of \eqref{symmetric-case-7}, we choose $a_{3}=0$, which provides transversal propagating modes, see the modes (\ref{symmetric-propagating-11B}), so that the correspondent energy velocities are given by
\begin{subequations}
\label{symmetric-case-11}
\begin{align}
V_{E}^{\pm} &= \frac{ S_{+}}{   U^{S}_{\pm} } \; ,
\quad \widetilde{V}_{E}^{\pm} = \frac{-{S}_{+}} { {U}^{S}_{\mp} } \; ,
\label{symmetric-case-12}
\end{align}
with
\begin{align}
U^{S}_{\pm} &=\frac{\mu\epsilon'}{2} + \frac{\mu\omega}{2} \frac{\partial \epsilon'}{\partial \omega} + \frac{1}{2}(S_{+}^{2}+S_{-}^{2}) + \nonumber \\
&\phantom{=}+ \frac{\mu^{2} |{\bf{a}}\times {\bf{c}}|^{2}}{32\omega^{2}} \cos^{2}\varphi \pm \frac{\mu S_{-} |{\bf{a}}\times {\bf{c}}|}{4\omega} \cos\varphi \; . \label{symmetric-case-11-1}
\end{align}
\end{subequations}
The general behavior of the energy velocity is depicted in Fig. \ref{plot-energy-velocity-symmetric-case-vplus-vtilplus} for $V_{E}^{+}$ and $\widetilde{V}_{E}^{+}$ and Fig. \ref{plot-energy-velocity-symmetric-case-vminus-vtilminus} for $V_{E}^{-}$ and $\widetilde{V}_{E}^{-}$. We observe a kind of mirror symmetry between the energy velocities related to the propagating modes of $n_{\pm}$ and the modes associated with $\tilde{n}_{\pm}$ (negative refraction) when $\varphi=0, \pi$. In fact, one notices that $V_{E}^{\pm}|_{\varphi=0} = V_{E}^{\mp}|_{\varphi=\pi}$ and $\widetilde{V}_{E}^{\pm}|_{\varphi=0} = \widetilde{V}_{E}^{\mp}|_{\varphi=\pi}$
Notice that if ${\bf a}$ is parallel (or anti-parallel) to ${\bf c}$, the contribution of the magnetic conductivity tensor is null
in the result (\ref{Eqsnsymm1}). Furthermore, the dispersion relation (\ref{Eqsnsymm1}) is equivalent to the dispersion equation
(\ref{realdispantysim}) of the antisymmetric case by replacing $\pm ({\bf a}\times{\bf c})/2 \rightarrow {\bf b}$, which assures similarity of results with the ones of the previous section.
By calculating the differential in relation to $k^{i}$-component in \eqref{Eqsnsymm1}, the corresponding group velocity is
\begin{align}
\label{symmetric-case-14}
{\bf v}_{g}^{\pm} = \frac{\partial \omega}{\partial {\bf k}} = \frac{ \displaystyle \frac{{\bf k}}{\mu \omega} \pm i \, \frac{({\bf a \times c})}{4\omega}}{ \displaystyle \left(\epsilon' + \frac{\omega}{2} \frac{ \partial \epsilon'}{\partial \omega} \right) + i \left(\epsilon'' + \frac{\omega}{2} \frac{ \partial \epsilon''}{\partial \omega} \right)} \; ,
\end{align}
which is valid for all frequency range, and for both solutions $k_{\pm}$ and $\tilde{k}_{\pm}$. Let us now writing the $k^{i}$-components as $k^{i} = k'^{i} + i k''^{i}$, then the group velocity can be rewritten as
\begin{align}
\label{symmetric-case-15}
{\bf v}_{g}^{\pm} &= \frac{ \displaystyle \frac{{\bf k}^{\prime}}{\mu\omega} + i \, \frac{\omega \epsilon'' \, \hat{{\bf k}} }{2 k'}} { \displaystyle \left( \epsilon' + \frac{\omega}{2} \frac{\partial \epsilon'}{\partial \omega} \right) + i \left(\epsilon'' + \frac{\omega}{2} \frac{\partial \epsilon''}{\partial \omega} \right) }  \; .
\end{align}

Similarly to the antisymmetric case of Sec.~\ref{antisymmetric-case-section}, we observe that for a non absorbing scenario, $\epsilon''=0$ and $|{\bf{a}}\times {\bf{c}}|\cos\varphi=0$, one has $V_{E}^{\pm} = v_{g}^{\pm}$.

\begin{figure}[H]
\begin{centering}
\includegraphics[scale=0.69]{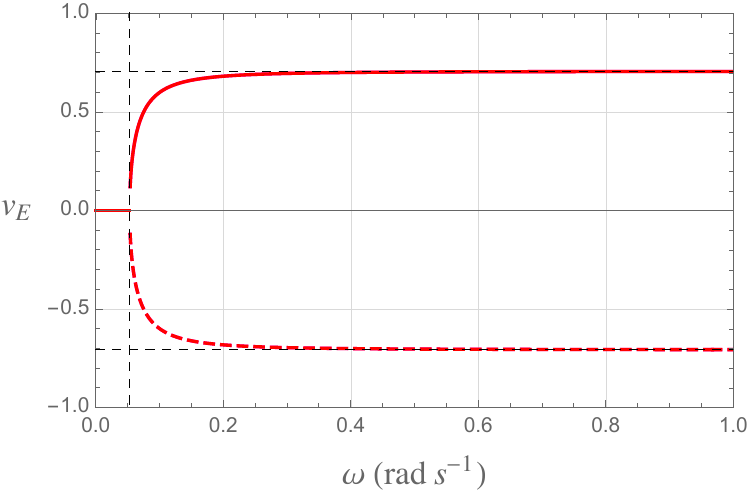}
\par\end{centering}
\caption{\label{plot-energy-velocity-symmetric-case-imaginary-part-epsilon-null} Energy velocity $V_{E}^{\pm}$ of \eqref{symmetric-case-12} for $\epsilon''=0$. Solid red line indicates $V_{E}^{\pm}$ for $\theta={0,\pi}$, while the dashed red curve represents $\widetilde{V}_{E}^{\pm}$ for $\theta={0,\pi}$. Here we have used $\mu=1$, $\epsilon'=2$, $|{\bf{a}}\times{\bf{c}}|=0.3$ $\mathrm{s}^{-1}$. The vertical dashed line indicates the frequency $\omega''$ of \eqref{symmetric-case-energy-velocity-observation-1}.}
\end{figure}

Furthermore, in the case of the magnetically induced absorption, $\epsilon''=0$ and $|{\bf{a}}\times {\bf{c}}|\cos\varphi \neq 0$, one also finds
\begin{align}
v_{g}^{\pm}=V_{E}^{\pm} = \frac{ {k'_{\pm}}/{\mu\omega}} { \displaystyle \epsilon' + \frac{\omega}{2} \frac{\partial \epsilon'}{\partial \omega} } \; ,  \label{symmetric-case-energy-velocity-extra-1}
\end{align}
where, in this case,
\begin{align}
k'_{\pm} = \omega S_{+}, \quad S_{\pm} = \frac{1}{\sqrt{2}} \sqrt{ |g(\omega)| \left(1 \pm \mathrm{sgn}[g(\omega)] \right) } \; , \label{symmetric-case-energy-velocity-extra-2}
\end{align}
such that,
\begin{align}
v_{g}^{\pm}=V_{E}^{\pm} = \frac{ \sqrt{ |g(\omega)| \left(1+\mathrm{sgn}[g(\omega)] \right)}}{ \displaystyle \mu \sqrt{2} \left( \epsilon'+ \frac{\omega}{2} \frac{ \partial \epsilon'}{\partial \omega} \right) } \; . \label{symmetric-case-energy-velocity-extra-3}
\end{align}
Again, we observe that
\begin{itemize}
\item For $g(\omega) >0$, one has $\mathrm{sgn} [g(\omega)]=1$, which yields
\begin{align}
v_{g}^{\pm}=V_{E}^{\pm} = \frac{ \sqrt{|g(\omega)| }}{ \displaystyle \mu \left(\epsilon' + \frac{\omega}{2} \frac{ \partial \epsilon'}{\partial \omega} \right) }
\; . \label{symmetric-case-energy-velocity-extra-4}
\end{align}
\item For $g(\omega) <0$, which occurs for $0<\omega<\omega''$, with
\begin{align}
	\omega''&= \sqrt{ \frac{ \mu |{\bf{a}}\times {\bf{c}}|^{2} \cos^{2}\varphi}{16 \epsilon'}} \; ,
	\label{symmetric-case-energy-velocity-observation-1}
\end{align}
there occurs $\mathrm{sgn} [g(\omega)]=-1$, leading to $S_{+}=0$, and
\begin{align}
v_{g}^{\pm}=V_{E}^{\pm} = 0 \; .
\label{symmetric-case-energy-velocity-extra-5}
\end{align}
\end{itemize}
This same analysis also holds for the energy velocity $\widetilde{V}_{E}^{\pm}$ associated with $\tilde{k}_{\pm}$, since $\tilde{k}_{\pm}'=-\omega S_{+}$. We illustrate the behavior of the energy velocities for $\epsilon''=0$ and $|{\bf{a}}\times~{\bf{c}}| \cos\varphi \neq 0$ in the Fig.~\ref{plot-energy-velocity-symmetric-case-imaginary-part-epsilon-null}, considering the parallel and antiparallel configurations.

\section{Final remarks}
\label{conclusions}

In this work, we have investigated a dielectric medium endowed with magnetic conductivity, concerning optical effects and electromagnetic signal propagation. Considering special cases for the magnetic conductivity, we have obtained the polarization of the collective electromagnetic propagating modes and their related refractive indices. We also have presented the solutions that describe negative refraction in each case. Furthermore, analyses of optical properties were carried out. For the isotropic magnetic conductivity of Sec.~\ref{particular2}, we determined the rotatory power, while the phase shift and absorption coefficients for the antisymmetric conductivity were found in Sec.~\ref{sec:nondiagonal-antisymmetric-configuration}. Finally, the symmetric conductivity was addressed in Sec.~\ref{sec:nondiagonal-symmetric-configuration}, where the phase shift and absorption difference (per unit length) between propagating modes were achieved.

The influence of crystal symmetries on the general magnetic conductivity tensor, $\sigma^{B}_{ij}$, was also addressed and is summarized in Tab.~\ref{symmetry-restrictions}, where one finds the maximal number of independent components it may have considering the restrictions brought about by the optical symmetry and related crystal system. The number of independent components of $\sigma^{B}_{ij}$ for each particular configuration (isotropic, antisymmetric, and symmetric) is in agreement with the expected number of components of general $\sigma^{B}_{ij}$ when the crystal symmetries restrictions are taken into account. Further details about the restrictions on the conductivity tensor are presented in Appendix \ref{symmetry-restriction-appendix}.

\vspace*{5pt}

\begin{widetext}
\hspace{2cm}  \begin{table}[H]
\caption{Symmetry restriction on arbitrary general magnetic conductivity tensor}
		\centering
		\begin{tabular}{ C{4cm}   C{8cm}  C{2.5cm}   }
			\toprule
		crystal system &  number of components of $\sigma^{B}_{ij}$ &optical symmetry   \\[1ex]
			\colrule \\[1ex]
		
cubic &   up to 7 in general; up to 5 when $\sigma^{B}_{ij}$ is symmetric, and up to 2 when  $\sigma^{B}_{ij}$ is antisymmetric &   isotropic 
\\
[0.6ex]
\\
	   trigonal, tetragonal, hexagonal	  &     up to   8 in general; up to 5 when it is symmetric, and up to 3 when it is antisymmetric   & uniaxial   \\
[0.6ex] \\
			orthorhombic, monoclinic, triclinic        & up to 9 in general  &    biaxial 
\\[0.6ex]
			\botrule
		\end{tabular}
		\label{symmetry-restrictions}
	\end{table}
\end{widetext}

In Sec.~\ref{section3}, we have investigated aspects regarding the propagation of electromagnetic energy. In doing so, we have obtained the energy velocity of the modes (related to positive and negative refraction). As expected, the energy velocity is negative for the negative refraction propagating modes. In this context, we have obtained some peculiar and surprising outcomes when comparing energy velocity and group velocity. For general non-absorbing media, it is known in the literature that energy velocity and group velocity are equivalent, whereas for absorbing media such equivalence is lost. However, our results revealed that, when the dispersive dielectric is endowed with magnetic conductivity, the latter statements are no longer valid. Indeed, for the isotropic conductivity of Sec.~\ref{energy-velocity-isotropic-case}, in the absence of absorption $(\epsilon''=0,\Sigma \neq 0)$, one finds that $v_{g}\neq v_{E}$, ``breaking" the equivalence between energy and group velocities for a real refractive index medium. Such a situation is represented in the first line, second column of Table \ref{table-velocity-comparison}.

For the anisotropic and non-diagonal conductivities, namely, the antisymmetric and symmetric cases of Secs.~\ref{antisymmetric-case-section} and \ref{symmetric-case-section}, respectively, the permittivity is non-reciprocal and also non-Hermitian, there occurring magnetically induced absorption, that is, the refractive indices become complex, even when $\epsilon''=0$, due to the non-null magnetic conductivity, as we can see in \eqref{antisymmetric-case-5} and \eqref{refractive-indices-symmetric-case-6-0}.  For both cases, it holds $v_{g}=V_{E}$ in spite of the associated complex refractive indices. In such cases, only when $\epsilon''\neq 0$ the energy velocity will be different from the group velocity. These cases are represented in the second and third lines of the second column of Table~\ref{table-velocity-comparison}.

These bold and surprising distinctions indicate that dispersive media endowed with magnetic conductivity have specific energy propagation properties, which seem to be not described by the usual formalism. All results regarding the group velocity and energy velocity in dispersive media endowed with magnetic conductivity are displayed in Table \ref{table-velocity-comparison}.

\begin{widetext}
\hspace{2cm}  \begin{table}[H]
\caption{Comparison of group and energy velocities in the cases of isotropic, antisymmetric, and symmetric magnetic conductivity tensors. Note that the unexpected results are the ones in the second column, with  $(\epsilon'' = 0,\hspace{0.2cm} \sigma^{B}_{ij} \neq 0)$.}
		\centering
		\begin{tabular}{ C{3.1cm}   C{3cm}  C{3cm}   C{3cm}       }
			\toprule
			&   $(\epsilon''\neq 0, \hspace{0.2cm} \sigma^{B}_{ij}\neq0)$ & $(\epsilon'' = 0,\hspace{0.2cm} \sigma^{B}_{ij} \neq 0)$  &   $(\epsilon''= 0, \hspace{0.2cm} \sigma^{B}_{ij}=0)$     \\[0.6ex]
			\colrule \\[0.1ex]
			$\sigma^{B}_{ij}=\Sigma\, \delta_{ij}$   &        $v_{g}\neq V_{E}$ $|$ \hspace{0.01cm} $n \in \mathbb{C}$            & $v_{g }\neq V_{E}$ $|$   \hspace{0.01cm} $n \in \mathbb{R}$    & $v_{g}= V_{E} $ $|$ \hspace{0.01cm} $n \in \mathbb{R}$
\\
[0.6ex]
			$\sigma^{B}_{ij}=\epsilon_{ijk}\,b_{k}$   &        $v_{g} \neq V_{E}$ $|$ \hspace{0.01cm} $n \in \mathbb{C}$            & $v_{g }= V_{E}$ $|$ \hspace{0.01cm} $n \in \mathbb{C}$       &   $v_{g}= V_{E} $  $|$ \hspace{0.01cm} $n \in \mathbb{R}$     \\
[0.6ex]
			$\sigma^{B}_{ij}=(a_{i}\,c_{j}+a_{j}\,c_{i})/2$         & $v_{g} \neq V_{E}$ $|$ \hspace{0.01cm} $n \in \mathbb{C}$ &    $v_{g} = V_{E}$  $|$ \hspace{0.01cm} $n \in \mathbb{C}$  & $v_{g} = V_{E} $  $|$ \hspace{0.01cm} $n \in \mathbb{R}$
\\[0.6ex]
			\botrule
		\end{tabular}
		\label{table-velocity-comparison}
	\end{table}
\end{widetext}

As a research possibility, we point that in the Weyl semimetals, the CME constitutive relation takes on the form, $J^{i} = \sigma^{CME}_{ij}\, E^{j}$, where $ \sigma^{CME}_{ij} = e^{2}/ (2\pi^{2}) \bar{\alpha} B^{i} B^{j}$, being proportional to $B^{2}$ for parallel electric and magnetic fields. This constitutive relation structure motivates investigations on dielectrics endowed with general nonlinear magnetic conductivities, as
 \begin{align}
 \sigma^{B}_{ij} &= \Sigma_{ij} + \alpha_{ijk}E^{k} + \eta_{ijk} B^{k} + \beta_{ijmn}E^{m}B^{n} +  \nonumber \\
 &\phantom{=} +\gamma_{ijmn} E^{m}E^{n} + \delta_{ijmn}B^{m}B^{n} + ... , \label{answer-8}
\end{align}
where the tensors of rank 3 and 4 parametrize the nonlinear response of the medium. Such scenarios, in which the permittivity and refractive indices depend on the fields, may be an interesting object of investigation in the future.

Furthermore, we highlight that the general formalism adopted in this work can be adapted to address bidimensional phenomena, such as the wave propagation in a surface, which is especially interesting in connection with plasmon and polariton excitations \cite{Mills} in the interface separation between two semi-infinite dielectric distinct media. Surface plasmon-polaritons can rise up as TM (transverse magnetic) modes when the permittivities of the two media are real and opposed in sign, that is, $\epsilon_{1}>0$, $\epsilon_{2}<0$, so that  $\epsilon_{1}\epsilon_{2}<0$. For an interface located in the plane $z=0$, the electric and magnetic fields may be written as
\begin{eqnarray}
	\mathbf{E}_{a} &=&(E_{ax},0,E_{az})\exp [i(k_{ax}x+k_{az}z-\omega t)], \\
	\mathbf{H}_{a} &=&(0,H_{ay},0)\exp [i(k_{ax}x+k_{az}z-\omega t)],
\end{eqnarray}
where $a=1,2$ designate the two semi-infinite media. Matching the boundary
conditions one obtains a superficial wave propagating along the x-axis ruled by
the dispersion relation
\begin{equation}
k_{x}=\frac{\omega }{c}\sqrt{\frac{\epsilon _{1}\epsilon _{2}}{\epsilon
		_{1}+\epsilon _{2}}}, \quad k_{az}=\frac{\omega }{c}\sqrt{\frac{\epsilon _{a}^{2}}{\epsilon
		_{1}+\epsilon _{2}}}.
\end{equation}
It is easy to show that $\epsilon _{1}>0,\epsilon _{2}<0,$ with $\left\vert
\epsilon _{2}\right\vert >\epsilon_{1}$ leads to a wave confined in the
plane $z=0$ with undamped propagation, that is, $k_{1z}=i\alpha \omega$,
$k_{2z}=-i\beta \omega$, $k_{x}=\gamma \omega$, being $\alpha,\beta,\gamma$
positive real numbers. In the case one of the media has absorption, the correspondent permittivity becomes complex and the wave propagation undergoes attenuation at the x-axis. Such a development is well known in the literature \cite{Raether,Yeh,Cottam,Cottam2,Agranovich3,Pitarke}. Surface plasmon-polaritons constitute an active line of research, having been recently considered in chiral/anomalous material \cite{Gorbar1} and in a strained slab of a Weyl semimetal with broken time-reversal symmetry \cite{Gorbar2}.
Further, as a proper continuation of the present work, we can examine the propagation of plasmon-polariton waves in the interface separating a usual dielectric medium, $\epsilon_{1}>0$, and an exotic dielectric endowed magnetic conductivity, whose permittivity $\overline{\epsilon}_{ij}(\omega)$, given in \eqref{permittivityCME}, can be diagonalized providing the principal values $\overline{\epsilon}_{11}$, $\overline{\epsilon}_{22}$, $\overline{\epsilon}_{33}$, depending on the magnetic conductivity parameters. The framework to address such an anisotropic system \cite{Raether,Yeh,Cottam,Cottam2,Agranovich3,Pitarke} is already applied for WSM \cite{Hofmann} and can be also adapted from the general procedure developed in the present work.

As already mentioned, a dielectric medium with magnetic current behaves as uniaxial or biaxial anisotropic matter (see the Appendix for more details). However, the question of comparing the present theoretical results with some experimental data is challenging, since so far experimental works involving an exotic dielectric medium in the presence of magnetic conductivity, $\sigma_{ij}^{B}$, considered as a property of the material, have not been reported. The absence of experimental works on this specific topic makes it difficult to perform some possible comparisons between our theoretical results and experimental data. Despite that, the order of magnitude of the magnetic conductivity $\sigma^{B}$ may be estimated approximately based on another investigation about magnetic currents (see Eq. (3) of \cite{Kaushik}), which is similar to our antisymmetric case, $\sigma_{ij}^{B}=\epsilon_{ijk}\,b_k$. The authors found that, for a class of Weyl semimetals of TaAs materials, currents of intensities of $0.75$ $\mu A$ (for type-I Weyl semimetal), and $2.5$ $\mu A$ (for type-II Weyl semimetal) are expected to occur when considering the numerical parameters in Table 1 on page 4 Ref. \cite{Kaushik}. Taking into consideration that our magnetic conductivity has the same order of magnitude as the conductivity expected in the transverse photocurrents in Weyl semimetals, we estimate the order of magnitude of $\sigma^{B}$ as follows. We write $i = J \, A = \sigma^{B} \, B \, A$, which yields
\begin{align}
\sigma^B \simeq \frac{i}{B} \, (m^{-2}) \; , \label{values-3}
\end{align}
where we are determining an approximate value for $\sigma^{B}$ per unit area $A$. Then, considering the numerical values used in Ref.~\cite{Kaushik}, we find $\sigma^{B} = 1.5 \times 10^{-6} \mathrm{A}  \, \mathrm{T}^{-1} \, \mathrm{m}^{-2}$ for type-I Weyl semimetal, and $\sigma^{B} =5 \times 10^{-6} \, \mathrm{A}\, \mathrm{T}^{-1} \, \mathrm{m}^{-2} \;$ for type-II Weyl semimetal.
These values seem to indicate that the magnetic conductivity considered in Ref.~\cite{Kaushik} has an order of magnitude $10^{-6}$  $\mathrm{A}  \, \mathrm{T}^{-1} \, \mathrm{m}^{-2}$, or, equivalently, $\sigma^{B} \sim 10^{-6} \, \Omega^{-1} \, s^{-1}$. Such an order of magnitude may be useful for experimental investigations.

As a final comment, we point out the additional possibility of addressing the relevant excitations of condensed matter realistic dielectric systems, described by the Drude frequency-dependent permittivities. These are much more involved functions in comparison with the simple choices for $\epsilon'$ and $\epsilon^{\prime\prime}$ adopted in the present work. This kind of study may be interesting for the optics of chiral dispersive dielectrics supporting magnetic conductivity and is being considered in a forthcoming investigation.

\begin{acknowledgments}
	
The authors express their gratitude to FAPEMA, CNPq, and CAPES (Brazilian research agencies) for their invaluable financial support. M.M.F. is supported by FAPEMA Universal/01187/18, CNPq/Produtividade 311220/2019-3 and CNPq/Universal/422527/2021-1. P.D.S.S is grateful to FAPEMA BPD-12562/22 and is currently supported by grant CNPq/PDJ 150584/23. Furthermore, we are indebted to CAPES/Finance Code 001 and FAPEMA/POS-GRAD-02575/21.

\end{acknowledgments}

\appendix

\section{\label{symmetry-restriction-appendix}Crystal symmetry restrictions on the magnetic conductivity tensor}

In this Appendix, we examine possible implications of certain crystal symmetries on the components of general magnetic conductivity tensor, supposing it general in principle. Afterward, we will consider the specific parametrizations for $\sigma^{B}_{ij}$ addressed in this work and the constraints on the crystal types.

\subsection{Materials symmetries constraints on arbitrary and general tensor $\sigma^{B}_{ij}$}
The permittivity effective tensor of the medium endowed with the magnetic conductivity $\sigma^{B}_{ij}$ is given in \eqref{permittivityCME},
\begin{align}
\label{crystal-symmetry-1}
\bar{\epsilon}_{ij} &= \epsilon \delta_{ij} - \frac{i}{\omega^{2}} \sigma^{B}_{ia} \epsilon_{abj} k_{b}, 
\end{align}
entirely written in the following matrix form:
\begin{widetext}
\begin{align}
\left[ \bar{\epsilon}_{ij} \right] &= \begin{bmatrix}
\epsilon - \frac{i}{\omega^{2} } \left( \sigma^{B}_{12} k_{3} - \sigma^{B}_{13} k_{2} \right) & & - \frac{i}{\omega^{2}} \left( - \sigma^{B}_{11} k_{3} + \sigma^{B}_{13} k_{1} \right)  &&  - \frac{i}{\omega^{2}} \left( \sigma^{B}_{11} k_{2} - \sigma^{B}_{12} k_{1} \right) \\
\\
- \frac{i}{\omega^{2}} \left( \sigma^{B}_{22} k_{3} - \sigma^{B}_{23} k _{2} \right)  && \epsilon - \frac{i}{\omega^{2}} \left( - \sigma^{B}_{21} k_{3} + \sigma^{B}_{23} k_{1} \right)  && - \frac{i}{\omega^{2}} \left( \sigma^{B}_{21} k_{2} - \sigma^{B}_{22} k_{1} \right) \\
\\
-\frac{i}{\omega^{2}} \left( \sigma^{B}_{32} k_{3} - \sigma^{B}_{33} k_{2} \right) && - \frac{i}{\omega^{2}} \left( - \sigma^{B}_{31} k_{3} + \sigma^{B}_{33} k_{1} \right) && \epsilon- \frac{i}{\omega^{2}} \left( \sigma^{B}_{31} k_{2} - \sigma^{B}_{32} k_{1} \right)  
\end{bmatrix}. \label{crystal-symmetry-2}
\end{align}
\end{widetext}
This matrix has three eigenvalues: $\epsilon$ and $\epsilon_{\pm}$, given below

\begin{subequations}
\label{crystal-symmetry-eigenvalues-generic}
\begin{align}
\epsilon_{\pm} &= \epsilon - \frac{i}{2\omega^{2}} \left( k_{1} g_{1} + k_{2} g_{2} + k_{3} g_{3}\right)  \pm \frac{1}{2\omega^{2}} \left[-k_{3}^{2} f_{33}  + \right. \nonumber \\
&\phantom{=} \left. - k_{2}^{2}f_{22} - k_{1}^{2}f_{11} + 2 H\right]^{1/2},   \label{crystal-symmetry-3}
\end{align}
with
\begin{align}
g_{1} &=  \sigma^{B}_{23} - \sigma^{B}_{32} , \\
g_{2} &=  \sigma^{B}_{31}-\sigma^{B}_{13} , \\
g_{3} &= \sigma^{B}_{12}-\sigma^{B}_{21}, \\ 
H&=k_{1}k_{2}f_{12}  + k_{1}k_{3}f_{13} + k_{2}k_{3} f_{23} , \\ 
f_{11} &= \left(\sigma^{B}_{23} + \sigma^{B}_{32} \right)^{2} - 4 \sigma^{B}_{22} \sigma^{B}_{33} , \\
f_{22} &= \left(\sigma^{B}_{31}+\sigma^{B}_{13} \right)^{2} - 4 \sigma^{B}_{11} \sigma^{B}_{33}, \\
f_{33} &= \left(\sigma^{B}_{12} + \sigma^{B}_{21} \right)^{2} - 4 \sigma^{B}_{11} \sigma^{B}_{22}, \\
f_{12} &=\left( \sigma^{B}_{13}+\sigma^{B}_{31} \right) \left( \sigma^{B}_{23}+\sigma^{B}_{32} \right) - 2 \sigma^{B}_{33} \left( \sigma^{B}_{12} + \sigma^{B}_{21} \right) , \\
f_{13}&= \left( \sigma^{B}_{12}+\sigma^{B}_{21} \right)  \left( \sigma^{B}_{23} +\sigma^{B}_{32}\right) - 2 \sigma^{B}_{22} \left(\sigma^{B}_{13} + \sigma_{31} \right) , \\
f_{23} &= \left(\sigma^{B}_{12} + \sigma^{B}_{21} \right)  \left( \sigma^{B}_{13}+\sigma^{B}_{31} \right) - 2 \sigma^{B}_{11} \left( \sigma^{B}_{23}+\sigma^{B}_{32} \right) 
\end{align}
\end{subequations}
having the diagonalized form,
\begin{eqnarray}
	\left[
	\begin{array}{ccc}
		\epsilon_{+} & 0 & 0 \\
		0 & \epsilon_{-} & 0 \\
		0 & 0 & \epsilon \\
	\end{array}
	\right] \;. \label{PTEG1}
\end{eqnarray}
In order to seek possible restrictions that certain crystal symmetries may impose on the general components of $\sigma^{B}_{ij}$, let us consider, without loss of generality, propagation along the $z$-axis. In this case, the three permittivity eigenvalues are  $\epsilon$ and,
\begin{align}
\epsilon_{\pm}&= \epsilon- \frac{i}{2\omega^{2}} k_{3} \left( \sigma^{B}_{12} - \sigma^{B}_{21} \right) + \nonumber \\
&\phantom{=} \pm \frac{i}{2\omega^{2}} k_{3} \sqrt{ \left( \sigma^{B}_{12} + \sigma^{B}_{21}\right)^{2} - 4 \sigma^{B}_{11} \sigma^{B}_{22} } . \label{crystal-symmetry-4}
\end{align}
In the following, we compare the effective diagonal permittivity (\ref{PTEG1}) of our dielectric supporting magnetic current with the one of cubic, uniaxial, and biaxial crystals. At a second moment, we try to associate it with the known crystal systems.  
\subsubsection{Cubic crystals}
In a cubic crystal, the permittivity tensor is totally isotropic, corresponding to three equal eigenvalues, that is,
\begin{align}
\left[ \epsilon\right]_{cubic} &= \begin{pmatrix}
a & 0 & 0 \\
0 & a & 0 \\
0 & 0 & a
\end{pmatrix} . \label{crystal-symmetry-5}
\end{align}
By requiring full correspondence between such a cubic crystal and the diagonalized permittivity (\ref{PTEG1}), it should hold $\epsilon=\epsilon_{+}=\epsilon_{-}$, providing the following relations:
  \begin{align}
 \sigma^{B}_{12} - \sigma^{B}_{21} &=0, \label{crystal-symmetry-6} \\
 (\sigma^{B}_{12} +\sigma^{B}_{21})^{2} - 4\sigma^{B}_{11} \sigma^{B}_{22} &=0 . \label{crystal-symmetry-7}
 \end{align}
Hence, the cubic crystal structure implies two 2 restrictions on the components of the tensor $\sigma^{B}_{ij}$, which is written as
\begin{align}
\left[ \sigma^{B}_{ij} \right] &= \begin{pmatrix}
\sigma^{B}_{11} && \sqrt{\sigma^{B}_{11} \sigma^{B}_{22} } && \sigma^{B}_{13} \\
\sqrt{\sigma^{B}_{11} \sigma^{B}_{22} } && \sigma^{B}_{22} &&  \sigma^{B}_{23} \\
\sigma^{B}_{31} && \sigma^{B}_{32} && \sigma^{B}_{33} 
\end{pmatrix} , \label{crystal-symmetry-8}
\end{align}
where we have used \eqref{crystal-symmetry-6} and \eqref{crystal-symmetry-7} to obtain $\sigma^{B}_{12} = \sqrt{\sigma^{B}_{11} \sigma^{B}_{22}} $. Notice that no constraints were obtained for the components $\sigma^{B}_{13}, \sigma^{B}_{23}, \sigma^{B}_{31}, \sigma^{B}_{32},$ and $\sigma^{B}_{33}$. Thus, the magnetic conductivity tensor $\sigma^{B}_{ij}$ can have, in principle, until 7 non-null components (in the principal axes optical system in which the permittivity (\ref{crystal-symmetry-2}) is diagonal).
A similar analysis proceeds in the case the propagation occurs along the x-axis, ${\bf{k}}=k \hat{\bf{x}}$. In this case, the eigenvalues of \eqref{crystal-symmetry-1} are $\epsilon$ and
\begin{align}
\epsilon_{\pm}^{x} &= \epsilon - \frac{i}{2\omega^{2}} k_{1} \left(\sigma^{B}_{23}-\sigma^{B}_{32} \right) + \nonumber \\
&\phantom{=} \pm \frac{i}{2\omega^{2}} k_{1}\sqrt{ \left(\sigma^{B}_{23} + \sigma^{B}_{32} \right)^{2} - 4 \sigma^{B}_{22} \sigma^{B}_{33} } .  \label{crystal-symmetry-9}
\end{align} 
Requiring that the effective permittivity (\ref{PTEG1}) be cubic, that is, $\epsilon=\epsilon_{+}^{x}=\epsilon_{-}^{x}$, one obtains, 
\begin{align}
\sigma^{B}_{23} - \sigma^{B}_{32} &=0, \label{crystal-symmetry-10} \\
 \left(\sigma^{B}_{23} + \sigma^{B}_{32} \right)^{2} - 4 \sigma^{B}_{22} \sigma^{B}_{33} &=0 . \label{crystal-symmetry-11}
 \end{align}
in such a way the conductivity tensor $\sigma^{B}_{ij}$ takes on the form,
\begin{align}
\left[\sigma^{B}_{ij} \right]_{x} &= \begin{pmatrix}
 \sigma^{B}_{11} && \sigma^{B}_{12} && \sigma^{B}_{13} \\
 \sigma^{B}_{21} && \sigma^{B}_{22} && \sqrt{ \sigma^{B}_{22} \sigma^{B}_{33}} \\
 \sigma^{B}_{31} && \sqrt{ \sigma^{B}_{22} \sigma^{B}_{33}} && \sigma^{B}_{33}
 \end{pmatrix} . \label{crystal-symmetry-12}
 \end{align}
Again, one notes that the cubic structure is compatible with until 7 non-null components for $\sigma^{B}_{ij}$, indicating that the elimination of two components does not depend on the propagation direction (in the principal axes system).
Furthermore, starting from the cubic compatible conductivity tensor (\ref{crystal-symmetry-12}), it holds:
\begin{itemize}
	\item In the case the conductivity is symmetric, up to five components may remain.
	\item In the case the conductivity is antisymmetric, up to two components may remain.	
\end{itemize}

\subsubsection{\label{uniaxial-section}Uniaxial crystals}
Uniaxial materials (tetragonal, hexagonal, and trigonal crystal systems) have the permittivity tensor with two equal eigenvalues \cite{Amnon}, that is
\begin{align}
\left[ \epsilon \right]_{uniaxial} &= \begin{pmatrix}
a & 0 &  0\\
0 & a & 0 \\
0 & 0 & b
\end{pmatrix} . \label{crystal-symmetry-13}
\end{align}
In this case, by requiring its correspondence with the effective diagonal permittivity (\ref{PTEG1}), it imposes that the eigenvalues $\epsilon_{\pm}$ of \eqref{crystal-symmetry-4} satisfy $\epsilon_{+} = \epsilon_{-}$, that is,
\begin{align}
\left(\sigma^{B}_{12}+\sigma^{B}_{21} \right)^{2} - 4\sigma^{B}_{11}\sigma^{B}_{22} &=0, \label{crystal-symmetry-14}
\end{align}
which implies
\begin{align}
	\left[ \sigma^{B}_{ij} \right] &= \begin{pmatrix}
		\sigma^{B}_{11} && \sigma^{B}_{12} && \sigma^{B}_{13} \\
		2 \sqrt{\sigma^{B}_{11} \sigma^{B}_{22} } - \sigma^{B}_{12} && \sigma^{B}_{22} && \sigma^{B}_{23} \\
		\sigma^{B}_{31} && \sigma^{B}_{32} && \sigma^{B}_{33} 
	\end{pmatrix}  . \label{crystal-symmetry-15}
\end{align}
Thus, the uniaxial optical structure imposes compulsorily only 1 restriction on the components of the tensor $\sigma^{B}_{ij}$ (having in mind the axis frame in which the effective permittivity becomes diagonal). Therefore, in this case, the magnetic conductivity tensor $\sigma^{B}_{ij}$ can have, in principle, until 8 non-null components (in the principal axes system). Further, considering the uniaxial compatible conductivity tensor (\ref{crystal-symmetry-15}), one has:
\begin{itemize}
	\item In the case the conductivity is symmetric, up to five components remain independent.
	\item In the case the conductivity is antisymmetric, up to three components remain independent.	
\end{itemize} 

\subsubsection{Biaxial crystals}
Biaxial materials (triclinic, monoclinic, and orthorhombic crystal systems) possess the permittivity tensor with 3 distinct eigenvalues \cite{Amnon}. Comparing such a scenario with the eigenvalues of \eqref{crystal-symmetry-4} (for ${\bf{k}}=k \hat{\bf{z}}$), we notice that there is no compulsory restriction on the components of $\sigma^{B}_{ij}$. Thus, in principle, the tensor $\sigma^{B}_{ij}$ can have until 9 non-null components in the proper axis frame.

\subsection{\label{Particular-cases-magnetic-conductivity}Specific cases of magnetic conductivity and relations to crystal systems}
We now examine the 3 distinct particular cases of magnetic conductivity examined in this work (isotropic, symmetric, and antisymmetric), tracing connections with crystal systems and optical symmetries.
\subsubsection{\label{isotropic-section}Isotropic case}
Considering the particular case where the magnetic conductivity is isotropic, $\sigma^{B}_{ij} = \Sigma \delta_{ij}$, the effective permittivity of \eqref{crystal-symmetry-1} becomes
\begin{align}
\bar{\epsilon}_{ij} &= \epsilon \delta_{ij} - \frac{i \Sigma}{\omega^{2}} \epsilon_{ibj} k_{b}, \label{crystal-symmetry-22}
\end{align}
whose matrix form is
\begin{align}
\left[ \bar{\epsilon}_{ij} \right] &= \begin{pmatrix}
\epsilon && \frac{i\Sigma}{\omega^{2}} k_{3} && - \frac{i \Sigma}{\omega^{2}} k_{2} \\
\\
-\frac{i \Sigma}{\omega^{2}} k_{3} && \epsilon && \frac{i\Sigma}{\omega^{2}} k_{1} \\
\\
\frac{i \Sigma}{\omega^{2}} k _{2} && - \frac{i \Sigma}{\omega^{2}} k_{1} && \epsilon
\end{pmatrix}, \label{crystal-symmetry-23}
\end{align}
with eigenvalues given by $\epsilon$ and $\epsilon_{\pm} = \epsilon \pm \frac{\Sigma}{\omega^{2}} k $, where $k=\sqrt{{\bf{k}}^{2}}$. In this case, one finds 3 distinct eigenvalues, being compatible with biaxial materials. Therefore, the isotropic conductivity can describe biaxial materials (triclinic, monoclinic, and orthorhombic crystal systems).
\subsubsection{Antisymmetric conductivity case}
In this scenario, the magnetic conductivity is parametrized by $\sigma^{B}_{ij} = \epsilon_{ijk} b_{k}$, yielding the effective permittivity 
\begin{align}
\bar{\epsilon}_{ij} &= \left[ \epsilon - \frac{i}{\omega^{2}} \left({\bf{k}}\cdot {\bf{b}} \right) \right] \delta_{ij} + \frac{i}{\omega^{2}} k_{i}b_{j}, \label{crystal-symmetry-25}
\end{align}
whose matrix form is
\begin{widetext}
\begin{align}
\left[ \bar{\epsilon}_{ij} \right] &= \begin{pmatrix}
\epsilon - \frac{i}{\omega^{2}} ({\bf{k}}\cdot {\bf{b}}) + \frac{i}{\omega^{2}} k_{1}b_{1} && \frac{i}{\omega^{2}} k_{1}b_{2} && \frac{i}{\omega^{2}} k_{1}b_{3}  \\
\\
\frac{i}{\omega^{2}} k_{2}b_{1} && \epsilon-\frac{i}{\omega^{2}} ({\bf{k}}\cdot {\bf{b}}) + \frac{i}{\omega^{2}} k_{2}b_{2} &&  \frac{i}{\omega^{2}} k_{2}b_{3} \\
\\
\frac{i}{\omega^{2}} k_{3}b_{1} && \frac{i}{\omega^{2}} k_{3}b_{2} && \epsilon - \frac{i}{\omega^{2}} ({\bf{k}}\cdot {\bf{b}}) + \frac{i}{\omega^{2}} k_{3}b_{3} 
\end{pmatrix} . \label{crystal-symmetry-26}
\end{align} 
\end{widetext}
The matrix (\ref{crystal-symmetry-23}) has two distinct eigenvalues $\epsilon$ and $\epsilon- \frac{i}{\omega^{2}} ({\bf{k}} \cdot {\bf{b}})$, which is compatible with uniaxial materials (embracing tetragonal, hexagonal, and trigonal crystal systems).  
It is also worthy to impose the skew-symmetry $\sigma^{B}_{ij}=-\sigma^{B}_{ji}$ in the previous generic eigenvalues of (\ref{crystal-symmetry-4}), implying
\begin{align}
\epsilon_{\pm}= \epsilon- \frac{i}{\omega^{2}} k_{3} \sigma^{B}_{12}, \label{crystalsymmetry-27}
\end{align}
since $\sigma^{B}_{12}=-\sigma^{B}_{22}$ and $\sigma^{B}_{11}=\sigma^{B}_{22}=\sigma^{B}_{33}=0$.
In this case, by comparing the conductivity $\sigma^{B}_{ij} = \epsilon_{ijk} b_{k}$ with the components of the maximal uniaxial form of \eqref{crystal-symmetry-15}, one obtains
\begin{align}
\sigma^{B}_{11}&=\sigma^{B}_{22}=\sigma^{B}_{33}=0, \label{crystal-symmetry-28} \\
\sigma^{B}_{12}&=-\sigma^{B}_{21}=b_{3}, \label{crystal-symmetry-29} \\
\sigma^{B}_{13} &= -\sigma^{B}_{31}=-b_{2}, \label{crystal-symmetry-30} \\
\sigma^{B}_{23} &= - \sigma^{B}_{32}=b_{1} . \label{crystal-symmetry-31}
\end{align}
in such a way the conductivity tensor presents only three non-null components, consistent with the maximal uniaxial matrix form \eqref{crystal-symmetry-15}.
\subsubsection{Non-diagonal symmetric case}
For the particular case where the magnetic conductivity is given by
\begin{align}
\sigma^{B}_{ij} &=\frac{1}{2} \left(a_{i}c_{j}+a_{j}c_{i} \right), \label{crystal-symmetry-32}
\end{align}
the effective permittivity becomes
\begin{align}
\bar{\epsilon}_{ij} &= \epsilon \delta_{ij} + \frac{i}{\omega^{2}} \left(a_{i}c_{n} + a_{n}c_{i} \right) \epsilon_{nbj}k_{b} , \label{crystal-symmetry-33}
\end{align}
whose matrix representation is
\begin{widetext}
\begin{align}
\left[\bar{\epsilon}_{ij} \right] &=
\left(
\begin{array}{ccc}
 \epsilon +\frac{i \left(k_2 \left(a_3 c_1+a_1 c_3\right)-k_3 \left(a_2 c_1+a_1 c_2\right)\right)}{2 \omega^{2} } & +\frac{i}{2 \omega } \left(2 a_1 c_1 k_3-k_1 \left(a_3 c_1+a_1 c_3\right)\right) & +\frac{i}{2 \omega^{2} } \left(k_1 \left(a_2 c_1+a_1 c_2\right)-2 a_1 c_1 k_2\right) \\
 +\frac{i}{2 \omega^{2} } \left(k_2 \left(a_3 c_2+a_2 c_3\right)-2 a_2 c_2 k_3\right) & \epsilon +\frac{i \left(k_3 \left(a_2 c_1+a_1 c_2\right)-k_1 \left(a_3 c_2+a_2 c_3\right)\right)}{2 \omega^{2} } & +\frac{i}{2 \omega^{2} } \left(2 a_2 c_2 k_1-k_2 \left(a_2 c_1+a_1 c_2\right)\right) \\
 +\frac{i}{2 \omega^{2} } \left(2 a_3 c_3 k_2-k_3 \left(a_3 c_2+a_2 c_3\right)\right) & +\frac{i}{2 \omega^{2} } \left(k_3 \left(a_3 c_1+a_1 c_3\right)-2 a_3 c_3 k_1\right) & \epsilon +\frac{i \left(k_1 \left(a_3 c_2+a_2 c_3\right)-k_2 \left(a_3 c_1+a_1 c_3\right)\right)}{2 \omega } \\
\end{array}
\right) . \label{crystal-symmetry-34}
\end{align}
\end{widetext}
In this case, there are 3 distinct eigenvalues for \eqref{crystal-symmetry-34}, namely, $\epsilon$ and also 
\begin{equation}
	\epsilon_{\pm} = \epsilon \pm \frac{i}{2\omega^{2}} \sqrt{ \left[ ({\bf{a}}\times {\bf{c}}) \cdot {\bf{k}} \right]^{2}}.\label{EIGANT2}
\end{equation}
Therefore, the non-diagonal symmetric conductivity \eqref{crystal-symmetry-32} can represent biaxial materials (monoclinic, triclinic, and orthorhombic crystal systems).
One also mentions that imposing the particular choice $\sigma^{B}_{ij}=\sigma^{B}_{ji}$ in our previous \eqref{crystal-symmetry-4} (for propagation along the $z$-axis), yields
\begin{align}
\epsilon_{\pm}&=\epsilon \pm \frac{i}{\omega^{2}} k_{3} \sqrt{ (\sigma^{B}_{12})^{2} - \sigma^{B}_{11} \sigma^{B}_{22} }. \label{crystal-symmetry-36}
\end{align}
Now, by replacing parametrization (\ref{crystal-symmetry-32}) in \eqref{crystal-symmetry-36}, one finds
\begin{align}
\epsilon_{\pm}= \epsilon \pm \frac{i}{2\omega^{2}} \sqrt{ k_{3}^{2} \, \left(a_{1}c_{2} - a_{2} c_{1} \right)^{2}}, \label{crystal-symmetry-39} 
\end{align}
which is the same result stemming from (\ref{EIGANT2}). This clearly illustrates that the isotropic and non-diagonal symmetric cases of $\sigma^{B}_{ij}$ are compatible with biaxial materials.

Table \ref{symmetry-and-crystal-systems} summarizes the correspondence between the conductivity tensor (isotropic, symmetric, and antisymmetric) with the optical symmetry and the possible crystal system.

\begin{widetext}
	\hspace{2cm}  \begin{table}[H]
		\caption{Optical symmetry of systems endowed with special cases of magnetic conductivity tensor}
		\centering
		\begin{tabular}{ C{3cm}   C{3cm}  C{3cm}   C{3cm}        }
			\toprule
			$\sigma^{B}_{ij}$ parametrization	&   number of independent components  & optical symmetry & crystal system \\[0.6ex]
			\colrule \\[0.1ex]
			$\sigma^{B}_{ij}=\Sigma\, \delta_{ij}$   &        1   & biaxial & orthorhombic, triclinic, monoclic
			\\
			[0.6ex]
			$\sigma^{B}_{ij}=\epsilon_{ijk}\,b_{k}$   &        3    & uniaxial    & trigonal, tetragonal, hexagonal \\
			[0.6ex]
			$\sigma^{B}_{ij}=(a_{i}\,c_{j}+a_{j}\,c_{i})/2$         & 6 &    biaxial & orthorhombic, triclinic, monoclic
			
			\\[0.6ex]
			\botrule
		\end{tabular}
		\label{symmetry-and-crystal-systems}
	\end{table}
\end{widetext}

\end{document}